\documentclass{aa}
\makeatletter
\let\linenumbers\relax
\let\linenumbers*\relax
\let\nolinenumbers\relax
\def\@LN{}
\def\@RN{}
\makeatother
\usepackage{graphicx}
\usepackage{txfonts}
\usepackage{lipsum}
\usepackage{subcaption}
\usepackage{lscape}
\usepackage{placeins}
\usepackage{graphicx}
\usepackage{amsmath}
\usepackage{amssymb}
\usepackage{multicol}
\usepackage{bm}
\usepackage{pdflscape}
\usepackage{subcaption}
\usepackage{enumitem}    
\usepackage[T1]{fontenc}
\usepackage{ae,aecompl}
\usepackage{newtxtext,newtxmath}
\usepackage{graphicx}
\usepackage{txfonts}
\usepackage{physics}
\usepackage{bm}
\usepackage[colorlinks=true, linkcolor=blue, citecolor=blue, urlcolor=blue]{hyperref}
\newcommand{\Msun}{\,\rm{M}_{\odot}}
\newcommand{\Zsun}{\,Z_{\odot}}

\begin{document}

   \title{Nitrogen enhancement of GN-z11 by metal pollution from supermassive stars}

   \subtitle{}


   \author{S. Ebihara\inst{1,2}
        \and M. Fujii\inst{1}
        \and T. Saitoh\inst{3}
        \and Y. Hirai\inst{4}
        \and H. Umeda\inst{1}
        \and Y. Isobe\inst{5,6,7}
        \and C. Nagele\inst{8}
        }

   \institute{Department of Astronomy, The University of Tokyo, 7-3-1 Hongo, Bunkyo-ku, Tokyo 113-0033, Japan
            \and Jodrell Bank Centre for Astrophysics, University of Manchester, Oxford Road, Manchester UK
            \and Department of Planetology, Graduate School of Science, Kobe University, 1-1 Rokkodai-cho, Nada-ku, Kobe, Hyogo 657-8501, Japan
            \and Department of Community Service and Science, Tohoku University of Community Service and Science, 3-5-1 Iimoriyama, Sakata, Yamagata 998-8580, Japan
            \and Kavli Institute for Cosmology, University of Cambridge, Madingley Road, Cambridge, CB3 0HA, UK
            \and Cavendish Laboratory, University of Cambridge, 19 JJ Thomson Avenue, Cambridge, CB3 0HE, UK
            \and Waseda Research Institute for Science and Engineering, Faculty of Science and Engineering, Waseda University, 3-4-1, Okubo, Shinjuku, Tokyo 169-8555, Japan
            \and Department of Physics and Astronomy, Johns Hopkins University, Baltimore, MD 21218, USA
            }

   \date{}

  \abstract
{Spectroscopic observations by the James Webb Space Telescope (JWST) have revealed young, compact, high-redshift ($z$) galaxies with high nitrogen-to-oxygen (N/O) ratios. GN-z11 at $z=10.6$ is one of these galaxies. }
{One possible scenario for such a high N/O ratio is pollution from supermassive stars (SMSs), from which stellar winds are expected to be nitrogen-rich. The abundance pattern is determined by both galaxy evolution and SMS pollution, but so far, simple one-zone models have been used. Using a galaxy formation simulation, we tested the SMS scenario.}
{We used a cosmological zoom-in simulation that includes chemical evolution driven by rotating massive stars (Wolf-Rayet stars), supernovae, and asymptotic giant branch stars.
As a post-process, we assumed the formation of an SMS with a mass between $10^3$ and $10^5 \Msun$ and investigated the contribution of its ejecta to the abundance pattern.}
{The N/O ratio was enhanced by the SMS ejecta, and the abundance pattern of GN-z11, including the carbon-to-oxygen and oxygen-to-hydrogen ratios, was reproduced by our SMS pollution model if the pollution mass fraction ranges within 10--30 per cent. This pollution fraction can be realized when the gas ionized by the SMS is polluted, and the gas density is $10^4$--$10^5$\,cm$^{-3}$ assuming a Str\"omgren sphere. We also compared the abundance pattern with those of other N/O-enhanced high-$z$ galaxies. Some of these galaxies can also be explained by SMS pollution.}
{}


   \maketitle


\section{Introduction}\label{sec:introduction}
The James Webb Space Telescope (JWST) has revealed the existence of nitrogen-rich galaxies at high redshift ($z$). A prime example of such galaxies is GN-z11. The galaxy was identified by the Hubble Space Telescope as having $z=11.09^{+0.08}_{-0.12}$ in \citet{Oesch2016}, and recent JWST observations updated its redshift to $z=10.6$ \citep{Bunker2023,Maiolino2024}. Several studies have analyzed the chemical abundance of GN-z11 and exhibited at least $\log(\mathrm{N/O})>-0.5$ \citep{Cameron2023,Isobe2023,Senchyna2024, Alvarez2025} from strong nitrogen lines in the rest-frame ultraviolet, regardless of whether a stellar or AGN radiation field is assumed. These nitrogen abundances are much higher than those of local galaxies \citep{Pilyugin2012, Izotov2006, Berg2016, Berg2019}, whose abundances correspond to the ejecta of core-collapse supernovae (CCSNe) \citep{Nomoto2013, Watanabe2024}. 
Such a high N/O ratio in some high-$z$ galaxies shows rather good agreement with that of globular clusters (GCs) in the Milky Way \citep{Charbonnel2023,Senchyna2024,Ji2025}. 

According to the traditional stellar evolution scenario, oxygen is mainly enriched by CCSNe, whereas carbon and nitrogen are enriched by stellar winds from asymptotic giant branch (AGB) stars. 
\citet{D'Antona2023} and \citet{McClymont2025} have suggested that AGB stars can be the origin of the N/O-enhanced gas of GN-z11. However, the main-sequence lifetimes of most progenitor stars (intermediate- or low-mass stars) are much longer than the age of high-$z$ galaxies, especially GN-z11, whose stellar age is estimated to be $24^{+20}_{-10}$ Myr \citep{Tacchella2023}. On the other hand, \citet{Rizzuti2025} assumed an SN-driven galactic wind model that selectively blows oxygen away after CCSNe.

A possible source of nitrogen that could have been ejected early on is Wolf-Rayet (WR) stars.  
In WR stars, the CNO cycle enriches nitrogen. The stellar wind from WR stars can also explain the N/O ratio of GN-z11 \citep{Isobe2023,Watanabe2024, 2024A&A...681A..30M},  
but a fine-tuned star formation history is required \citep{Cameron2023, 2024A&A...681A..30M}. 
Thus, \citet{Gunawardhana2025} claimed that the WR-only scenario can explain the carbon enrichment of GN-z11 but not its nitrogen excess.
Tidal disruption events (TDEs) are also a possible scenario \citep{Isobe2023}. This scenario requires a supermassive black hole. GN-z11 may host an active galactic nucleus (AGN), but this remains under debate \citep{Bunker2024, Maiolino2024, 2024A&A...681A..30M}.

Another possible source is very massive stars (VMSs) and supermassive stars (SMSs). VMSs and SMSs are stars with masses of $10^3$--$10^4$ and $>10^4\Msun$, respectively. 
So far, no such massive stars have been observed, but they can be formed as Population (Pop) III stars \citep{Latif2014,Chon2018,Latif2021,Hirano2021} or via runaway collisions of stars \citep{Devecchi2009,Sakurai2017,Fujii2024,Rantala2024}.
Note that Pop III stars are thought to form when the gas metallicity is below the critical threshold $Z_{\rm crit}=10^{-4}\,Z_\odot$, which marks the transition from Pop III to Pop II \citep{2001MNRAS.328..969B, 2002A&A...382...28S}. 
In contrast, the runaway collision scenario remains applicable regardless of the gas metallicity.
Although the evolution and yield of V/SMS have been little studied, several studies have proposed that V/SMS can also explain the N/O ratio of GN-z11 and other objects with V/SMS of stars with $Z=0.1\ Z_{\odot}$ and Pop III stars, respectively \citep{Charbonnel2023,Nagele2023,Nandal2024,Marques-Chaves2024}.
Hereafter, we refer to this V/SMS scenario simply as the SMS scenario. 

The formation of SMSs during the formation of GN-z11-like galaxies would be expected from their compactness. The half-mass (half-light) radius of GN-z11 is measured to be only 200\,pc \citep{Tacchella2023}. From the mass of $10^9\Msun$, the half-mass density of GN-z11 is estimated to be $\sim 100\Msun$\,pc$^{-3}$. Assuming a power-law density distribution with a power of $-2$, we expect $10^6 \Msun$\,pc$^{-3}$ at $\sim 1$\,pc. In star cluster formation simulations, such a high density is shown to result in runaway collision of stars \citep{PortegiesZwart2002} and the formation of V/SMSs \citep{Fujii2024,Rantala2024}. In addition, the N/O-enhanced population is known in GCs of the Milky Way \citep{Charbonnel2023}, and V/SMSs have been suggested for GCs \citep{Gieles2018}. Thus, SMS formation in GN-z11-like galaxies seems to be a natural consequence. 

Most previous numerical studies of N-enrichment due to SMSs have used one-zone models \citep{Fukushima2025, Watanabe2024}, and a few studies have used non one-zone multi-zone models \citep{Kobayashi2024,2026ApJ...997..309S}. However, real galaxies formed through more complex dynamical and chemical evolution. 
\citet{Saitoh2025} performed a numerical simulation of galaxy formation similar to GN-z11. The simulation showed that N/O is enriched during the first 10--20 Myr after a bursty star formation due to the stellar winds from rotating massive stars, although the N/O ratio was slightly lower than that of the observational lower limit of GN-z11. 
Since their simulation did not include SMSs, it remains unclear whether SMSs are responsible for the observed level of N-enrichment in realistic environments of galaxy formation.

In this paper, we test the SMS scenario using results from cosmological simulation presented in \citet{Saitoh2025}.
From the dynamical evolution of the galaxy, we can assume the formation of SMSs as a post-process.
In contrast to previous studies using one-zone models \citep{Fukushima2025, Watanabe2024}, our 3D model can trace a spatial distribution of GN-z11-like galaxies, thereby providing a more realistic chemical and dynamical evolution.

This paper consists of the following sections. In Section~\ref{sec:Method}, the calculation code and the setup of our galaxy formation simulation are described. 
Section~\ref{sec:Results} presents the results of our simulation and shows that our SMS scenario can reproduce chemical abundances similar to those of GN-z11. 
In Section~\ref{sec:Discussion}, we discuss whether the SMS scenario can explain other N/O-enhanced galaxies. 
We summarize our research in Section~\ref{sec:Conclusion}.

\section{Method}\label{sec:Method}

\subsection{Numerical simulation} \label{sec:code}
We adopt the simulation results presented in \citet{Saitoh2025}. 
In the following, we briefly summarize the simulation.
The simulation was performed using a zoom-in technique for a halo taken from a cosmological structure formation simulation.
The target halo is the most massive in a $100\,h^{-1}$Mpc$^3$ volume at $z\sim10$. Its mass is comparable to the expected mass of GN-z11 \citep{Scholtz2024}. Using \textsc{MUSIC} \citep{Hahn2011}, the initial condition was generated assuming standard cosmological parameters with \citet{Planck2020} ($H_0 = 67.32$, $\Omega_{\mathrm{M}} = 0.3158$, $\Omega_{\mathrm{\Lambda}} = 0.6842$, $\Omega_{\mathrm{b}} = 0.04938$, and $\sigma_8 = 0.812$). The mass resolutions of $4628\Msun$ and $24972\Msun$ for baryons and dark matter, respectively, for the zoomed-in region.
The softening lengths are $\epsilon_{\mathrm{baryon}}=5.7\,\mathrm{pc}$ and $\epsilon_{\mathrm{DM}}=12.6\,\mathrm{pc}$ for baryons and dark matter, respectively. The boundary particles have larger masses and softening lengths than those in the zoomed-in region.

The cosmological zoom-in simulation was performed using \textsc{ASURA} \citep{Saitoh2008,Saitoh2009}, which is a parallel $N$-body/smoothed particle hydrodynamics (SPH) code that solves gravitational interactions with the Tree method \citep{Barnes1986} and models hydrodynamical interactions with the density-independent formulation of SPH \citep{Saitoh2013, Saitoh2016}. 
\textsc{Grackle} \citep{Smith2017} is used to evaluate radiative cooling and heating of gas, directly solving the evolution of 12 species while using pre-computed tables for metal cooling and heating constructed by \textsc{Cloudy} \citep{Ferland2013}.
The ultraviolet background of \citep{HaardtMadau2012} and the Lyman-Werner background of \citep{Incatasciato2023} are included as external radiation fields.

Star formation is implemented based on a simple stellar population (SSP) approximation. A gas particle forms a star particle with an efficiency of $C_{*} = 0.5$ when it satisfies the following conditions: (i) high density ($n_{\mathrm {H}} > 10^3~ {\mathrm {cm}}^{-3}$), (ii) low temperature ($T <100$~K), and (iii) converging flow ($\div \bm v <0$). 
For higher-metallicity stars ($Z > 10^{-5}~\Zsun$), SSP particles follow the Chabrier initial mass function (IMF) \citep{Chabrier2003} with a mass range of $0.1$--$100~\Msun$, while for low-metallicity stars ($Z < 10^{-5}~\Zsun$), the Susa IMF \citep{Susa2014} with a mass range of $0.7$--$300~\Msun$ is used.

Stellar evolution and feedback are handled with \textsc{CELib} \citep{Saitoh2017}, including four chemical enrichment processes: stellar winds from rotating massive stars, CCSNe, AGB mass loss, and Type Ia SNe. Yield tables from \citet{Limongi2018}, \citet{Nomoto2013}, the FRUITY database \citep{FRUITY2011, FRUITY2014, FRUITY2015, FRUITY2016}, and \citet{Seitenzahl2013} are used for these processes. The momentum feedback model of \citet{Hopkins2018} is applied, with CCSNe releasing $10^{51}$--$10^{52}\,{\mathrm {erg}}$.
Metal diffusion driven by turbulence \citep{Shen2010} is taken into account, with a diffusion coefficient of 0.01 \citep{Hirai2017}. 

The output interval in the original \citet{Saitoh2025} simulation is insufficient to assess the role of SMSs. Hence, we run a new simulation starting from the output at $z \sim 11.2$ (corresponding to a cosmic age of $402~\mathrm{Myr}$) with an output interval of $1.38~\mathrm{Myr}$. While the overall evolution remains consistent with the original simulation, we observe slight differences, e.g., in the star formation rate and the absolute values of the chemical composition, due to numerical fluctuations.

\subsection{Supermassive star models}\label{sec:SMS_models}

In this paper, we consider the pollution from a single SMS that forms the galactic center. Thus, we adopt the stellar evolution and hydrodynamic simulations of a single SMS model by \citet{Nagele2023}. This work considered the SMS with $Z=0.1\,Z_\odot$. This metallicity is roughly consistent with that of GN-z11.
They first performed stellar evolution simulations for $10^3$, $10^4$, $5\times10^4$, and $10^5$\,$\Msun$ SMS models using the post-Newtonian stellar evolution code HOSHI \citep{Takahashi2016, Takahashi2018, Takahashi2019, Yoshida2019, Nagele2020}. These models included mass loss due to line-driven winds, and the enriched material in these winds is partly responsible for the chemical yields. Strictly speaking, a star with $10^3$\,$\Msun$ probably should be classified as a VMS, but for simplicity, we will refer to all four models as SMSs. The HOSHI code solves the hydrodynamic equations and heat diffusion within a 1D stellar structure using the Henyey method. HOSHI includes a nuclear reaction network of 52 isotopes (including the CNO cycle), neutrino cooling, convection, mass loss, and rotation.

After the stellar evolution simulations were completed, they performed a GR stability analysis \citep{Nagele2022MNRAS.517.1584N,Nagele2023a} on each time snapshot. SMSs are radiation-dominated and, as such, they are vulnerable to collapse induced by the general relativistic radial instability \citep{chandrasekhar1964}. This is not necessarily the only instability that can destabilize these stars, but it is thought to be the first one to activate during their evolution. The stability analysis developed in \citet{Nagele2022MNRAS.517.1584N} involves solving for the oscillation frequency of a perturbed SMS, which can be done for a set of hydrostatic stellar properties. If the frequency squared is less than zero, the perturbation will induce exponential motion as opposed to sinusoidal, and the star will collapse. 

\begin{figure}
    \centering
    \includegraphics[width=0.8\columnwidth]{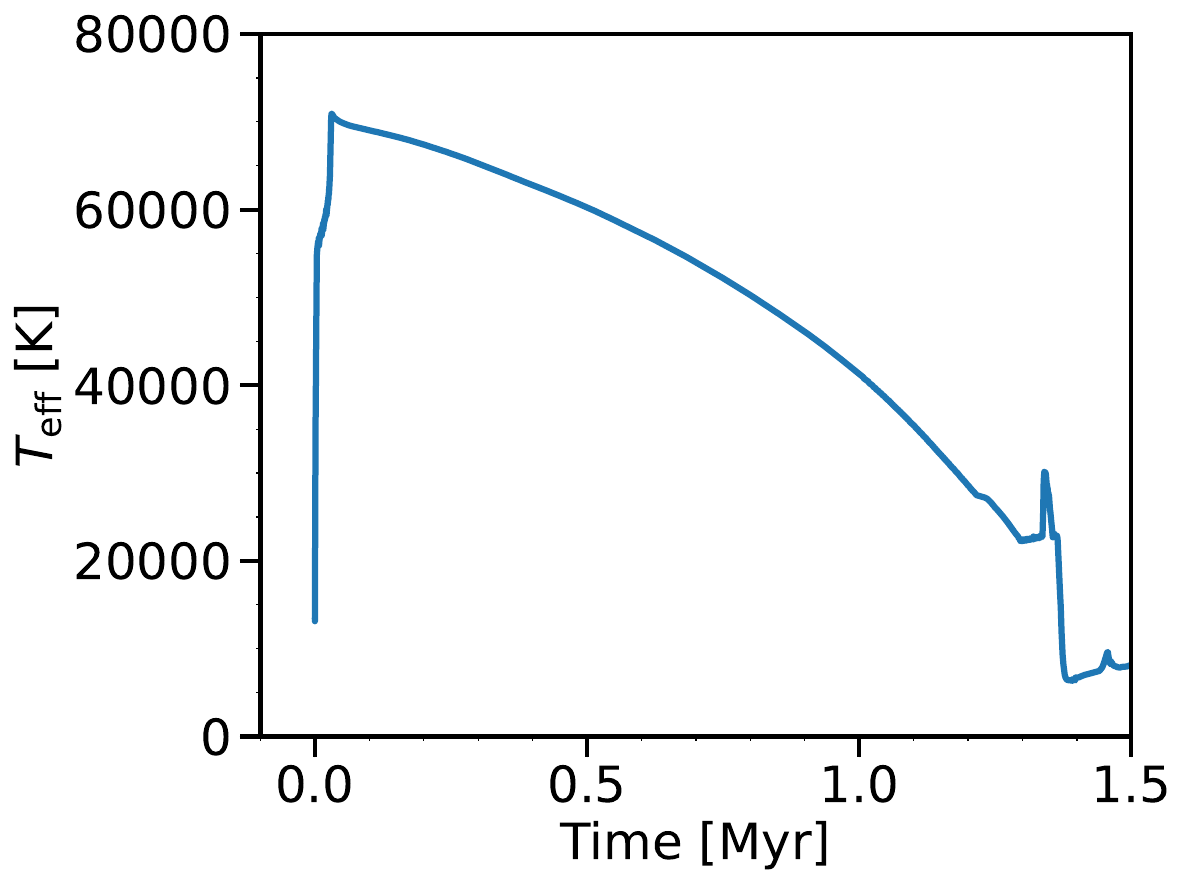}
    \caption{The effective temperature evolution of $10^4$\,$\Msun$ SMS model performed in \citet{Nagele2023}.
    }
    \label{fig:T_eff_evolution}
\end{figure}

This point of collapse is obviously the end of the star's lifetime, but in order to define a stellar age, we need the starting point of the star's lifetime. There is no obvious zero-age main-sequence point for massive stars. Therefore, here we define the start of the SMS life as occurring at $\log T_\mathrm{c}=7.5$. Given this definition, the stellar lifetimes are 1.7, 1.6, 1.3, 0.65 Myr for $M=10^3,\;10^4,\;5\times 10^4,\;10^5\;\Msun$, respectively. Fig. \ref{fig:T_eff_evolution} shows the effective temperature as a function of time after this initial time for the $10^4$\,$\Msun$ SMS model. The maximum effective temperature was $T_{\mathrm{eff}} = 7.09\times10^4$\,K, and the radius at that moment was $124$\,$R_\odot$. Unlike accreting SMSs \citep{Haemmerle2019A&A...632L...2H,Nagele2024}, the models considered here are relatively compact with higher surface temperatures. 

Once the GR radial instability is found in a given evolutionary snapshot, the authors transferred that snapshot into a Lagrangian 1D GR hydrodynamics code coupled to a large nuclear network \citep{Yamada1997,Takahashi2016,Nagele2020,Nagele2022MNRAS.517.1584N,Nagele2023a}. Running this code determined whether the unstable star would collapse or explode due to nuclear burning (in this case, the CNO cycles and proton captures of other light elements). The yields discussed in this paper include the material in the winds from the evolutionary calculation, as well as any material that is unbound during a pulsation or explosion. 

During the evolutionary simulation, the two heavier models, $5\times10^4$ and $10^5$\,$\Msun$, became GR unstable around the helium-burning phase. The two lower-mass models, $10^3$ and $10^4$\,$\Msun$, remained stable during helium-burning and finally formed carbon-oxygen cores. Extreme convection dredged up helium, carbon, and oxygen from the core, so the surfaces of these stars (and hence the composition of the winds) contained super-solar abundances of carbon and oxygen, in contrast to the heavier models, which were more nitrogen-dominated. 
Thus, the star's stability at different nuclear-burning stages affects the character of the \textit{total} yields it produces.

In Table~\ref{tab:Nagele_2023_table}, we summarize the yields of each SMS model, as given in Table~1 of \citet{Nagele2023}. Comparing the four models, the $10^3$, $10^4$, and $5\times 10^4\;\Msun$ models have higher oxygen abundances than carbon and nitrogen abundances. For these three models, the nitrogen-to-oxygen ratio is highest at $5\times 10^4\;\Msun$, which does not reach the late helium-burning phase. 
Nitrogen is also higher than carbon in all three models. The $10^5\;\Msun$ model has more carbon and nitrogen than oxygen, primarily due to the thermonuclear explosion, which produces large amounts of carbon and nitrogen (Table~\ref{tab:Nagele_2023_table}). The overall amount of CNO elements is similar to the lower mass models; however, due to this model's collapse earlier in its evolution, the total mass lost to winds is lower.

\begin{table*}
 \caption{Yields of SMSs.
 }
 \label{tab:Nagele_2023_table}
 \begin{center}
 \begin{tabular*}{1.6\columnwidth}{@{}cccccccc@{}}
  \hline
  \shortstack{$M_{\rm{SMS}}$\\$[\Msun]$} & \shortstack{Enrichment source\\$\;$} &  \shortstack{$E_{\rm explosion}$\\$[$erg$]$}&\shortstack{H\\$[\Msun]$} & \shortstack{He\\$[\Msun]$} & \shortstack{C\\$[\Msun]$} & \shortstack{N\\$[\Msun]$} & \shortstack{O\\ $[\Msun]$}\\
  \hline
  $10^3$ & wind &---&205 & 334 & 9.91 & 17 & 27\\[2pt]
  $10^4$ & wind&---&$1.65\times10^3$ & $2.73\times10^3$ & 1.81 & 75.8 & 133\\[2pt]
  $5\times10^4$ &wind&--- & $5.06\times10^3$ & $2.7\times10^3$ & 0.142 & 6.69 & 0.412\\[2pt]
  $10^5$ &wind/explosion&$7.97\times10^{54}$& $4.19\times10^4$ & $5.8\times10^4$ & 5.63 & 81.4 & 0.884\\[2pt]
  \hline
 \end{tabular*}
  \tablefoot{Stellar masses, properties, and yields (hydrogen, helium, carbon, nitrogen, and oxygen) of SMSs from Table 1 in \citet{Nagele2023}.
 }
 \end{center}
\end{table*}

\section{Results} \label{sec:Results}

\subsection{Galaxy formation}\label{sec:galaxy_formation}

\begin{figure*}[h]
    \centering
    \includegraphics[width=1.8\columnwidth]{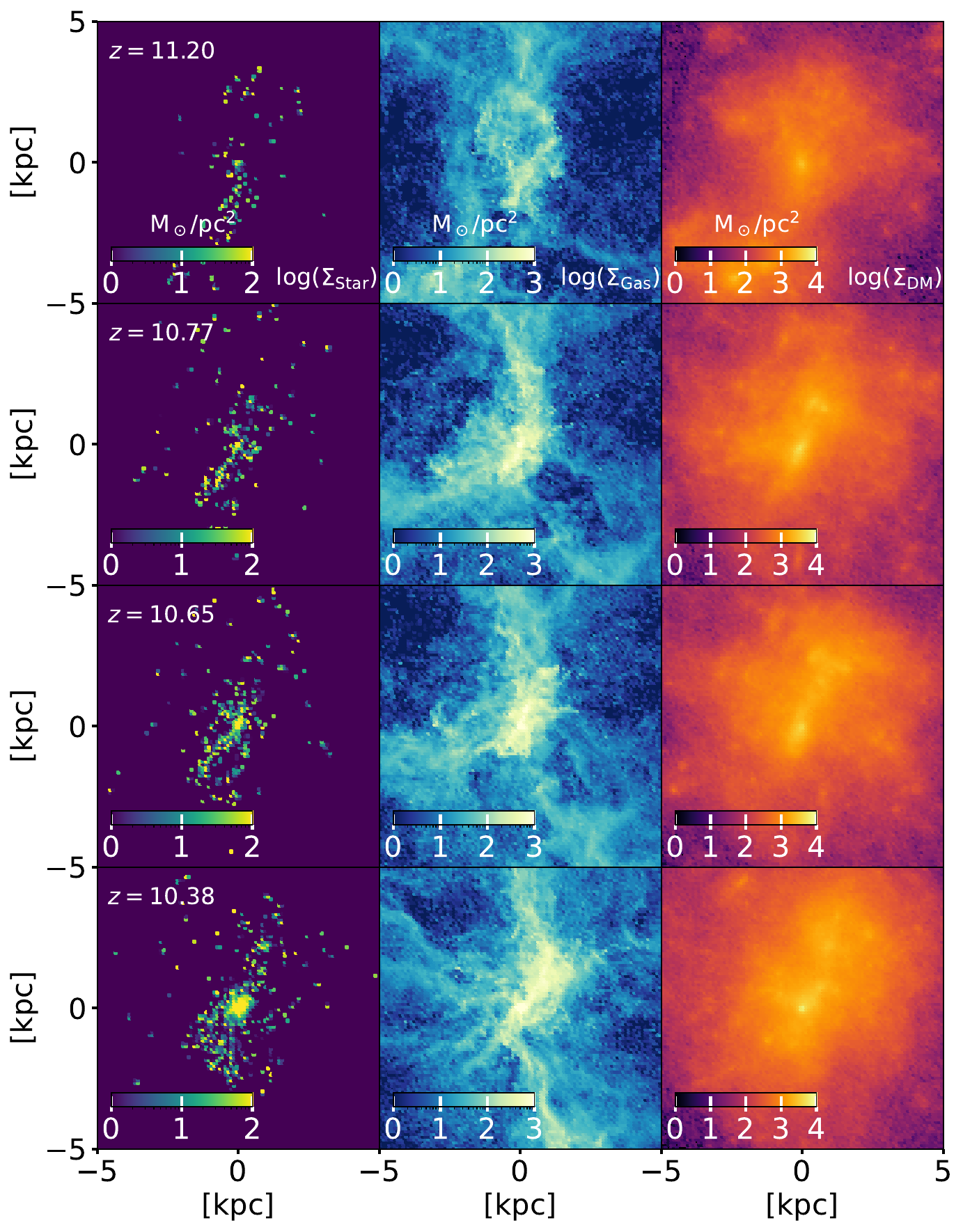}
    \caption{Projected surface densities of stars (left), gas (middle), and dark matter (right) at $z=11.20$, 10.77, 10.65, and 10.38 from top to the bottom.  
    }
    \label{fig:heatmap_evolution}
\end{figure*}

We present the evolution of the projected surface density of each component (stars, gas, and dark matter) in Fig.~\ref{fig:heatmap_evolution}. 
As shown in these snapshots, the galaxy evolved via mergers of three dark matter clumps between $z=11.20$ and $z=10.77$.
One of the merging dark matter clumps brought gas into the main halo (see the clump at lower left in the $\Sigma_{\mathrm{DM}}$ figure at $z=11.20$). Such gas accretion triggered star formation.

\begin{figure}
    \centering
    \includegraphics[width=0.9\columnwidth]{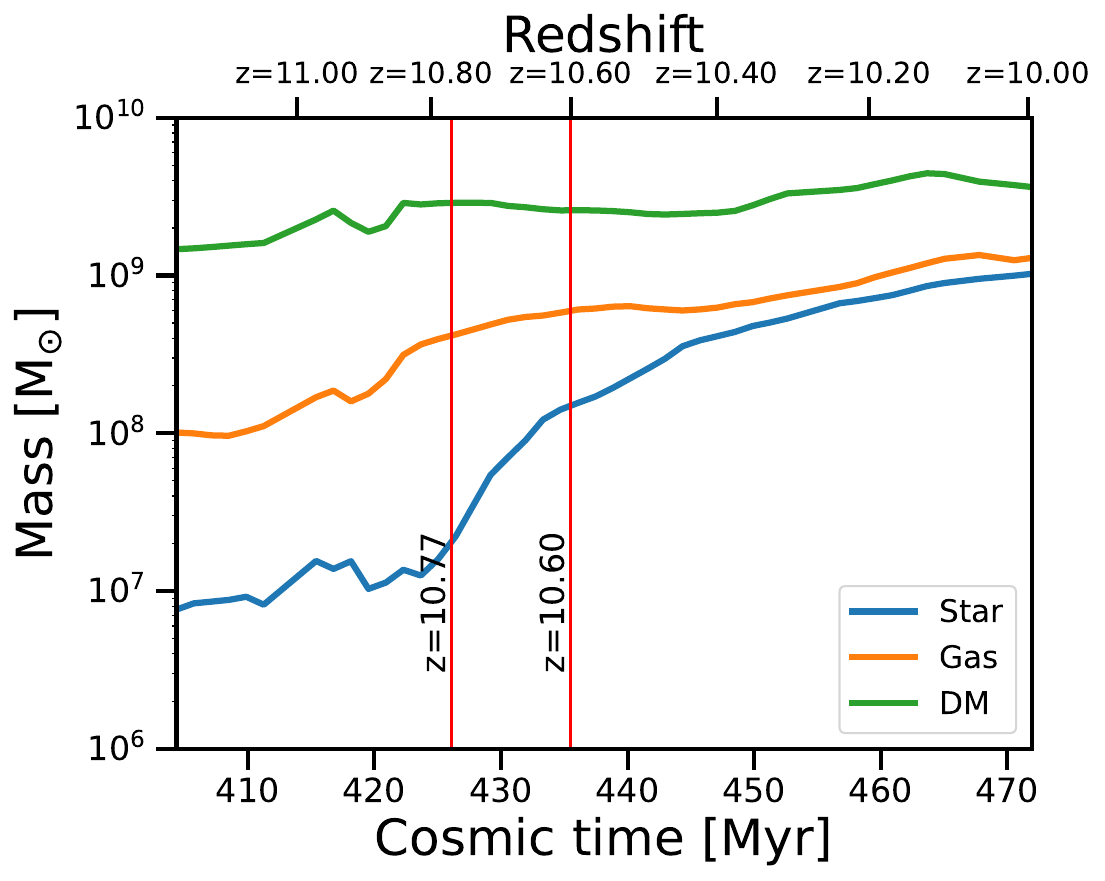}
    \caption{Mass evolution of stars (blue), gas (orange), and dark matter (green) within 1\,kpc from the galactic center. Redshifts of each snapshot are noted above the upper $x$ axis. The red vertical lines indicate the moment of $z=10.77$ and $z=10.60$.}
    \label{fig:3type_mass_evolution}
\end{figure}

The time evolution of the mass of our simulated galaxy is shown in Fig.~\ref{fig:3type_mass_evolution}. Here, we define the mass within 1\,kpc of the galactic center as the galaxy's mass. Note that the galactic center was determined using \textsc{Amiga's Halo Finder} \citep[AHF; ][]{Steffen2009}, considering the density of dark matter and baryons.
The dark matter mass was almost constant ($10^{9.5}$\,$\Msun$) between 410 and 470 Myr. On the other hand, the gas mass rapidly increased from $\sim 420$\,Myr. As a result, the stellar mass also rapidly increased after $\sim420$\,Myr ($z=10.80$). At this moment, the mergers of the three dark-matter clumps occurred (see Fig.~\ref{fig:heatmap_evolution}). The stellar mass reached $\sim10^8\ \Msun$ until $z=10.60$. 
This value is still slightly smaller than the observationally estimated stellar mass of GN-z11; $\log(M_*/\Msun)=8.7\pm0.06$ \citep{Bunker2024} or $\log(M_*/\Msun)=9.1^{+0.3}_{-0.4}$ \citep{Tacchella2023}.

The star formation slightly slows down after $z=10.60$ but continues until $470$\,Myr. The stellar mass finally reached $10^9\Msun$, comparable to GN-z11. 
Although the stellar mass at $z=10.60$ is slightly lower than that of GN-z11, we assume that an SMS formed around this time, as it is just after the rapid increase in stellar mass. Previous numerical studies on star cluster formation suggested that an SMS can be formed via runaway collisions during a rapid star formation of the system \citep{Fujii2024,Rantala2024}.

\begin{figure}
    \centering
    \includegraphics[width=1\columnwidth]{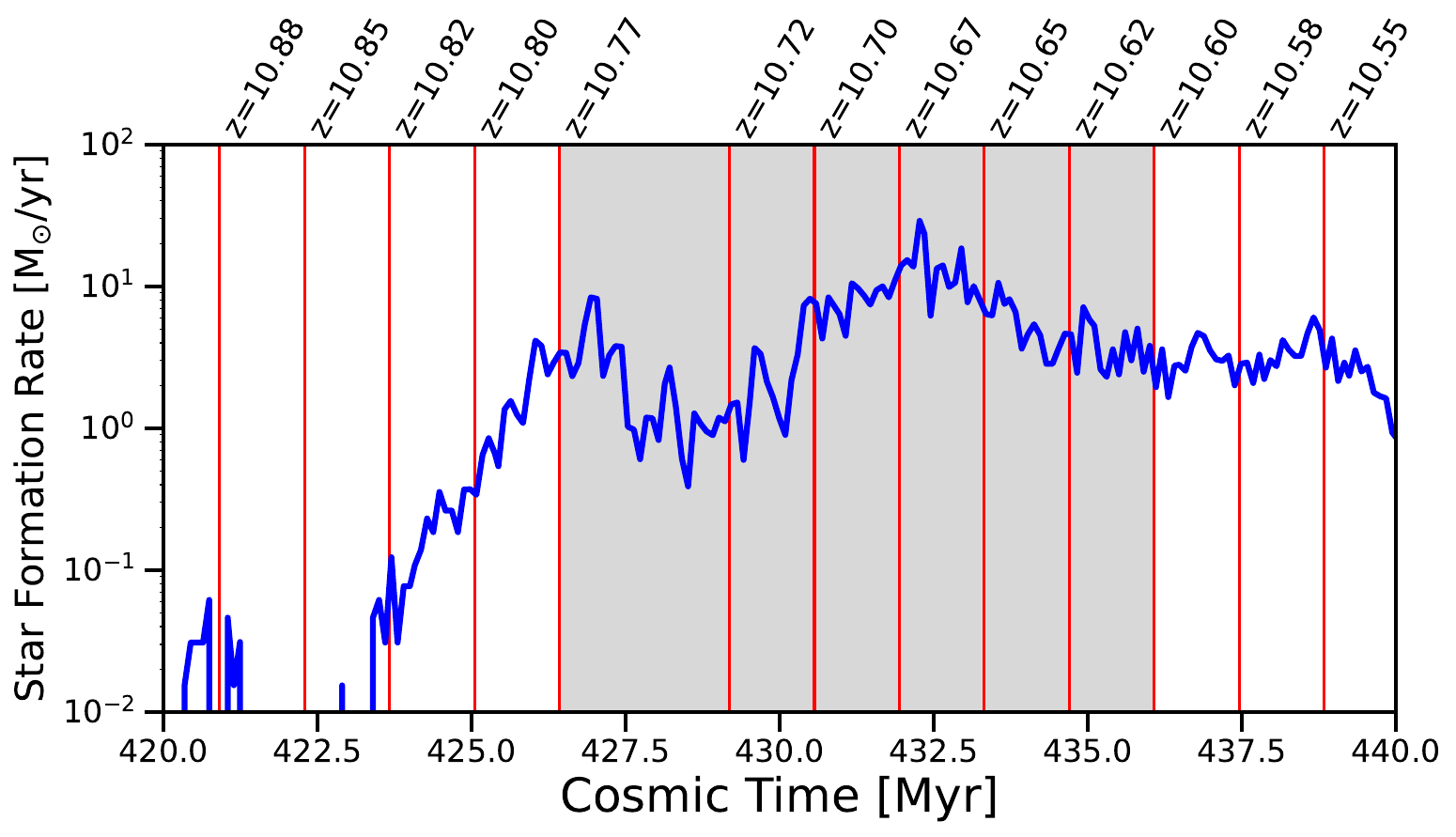}
    \caption{Time evolution of the star formation rate in the central region (within 10\,pc) of our simulated galaxy. 
    Red vertical lines indicate the snapshot times. Redshifts of each snapshot are noted above the upper $x$ axis. The shaded region indicates the time range between $z=10.77$ and $10.60$.}
    \label{fig:long_SFR_connected}
\end{figure}

We also measured the star formation rate (SFR) of the central region (10\,pc) of this galaxy because stellar collisions are expected to occur at the galactic center. 
For each snapshot, we identified star particles within 10\,pc of the galactic center. 
From the formation-epoch data for each star particle, we calculated the SFR in bins of $10^4$\,yr between the target snapshot and the previous one. Repeating this for each snapshot, we constructed the SFR history, as shown in Fig.~\ref{fig:long_SFR_connected}.
The SFR was fluctuating around 10\,$\Msun$\,$\mathrm{yr}^{-1}$ between $z=10.77$ and $z=10.60$. The SFR reached the peak, $>10$\,$\Msun$\,$\mathrm{yr}^{-1}$, between $z=10.67$ and $z=10.65$. In \citet{Fujii2024}, the SFRs of their star clusters were typically $\sim 1$\,$\Msun$\,$\mathrm{yr}^{-1}$ and reached $\sim 10$\,$\Msun$\,$\mathrm{yr}^{-1}$ for the most massive ones, and their star clusters resulted in the formation of stars with $>10^3 \Msun$. 
Given this discussion, we assume that a single SMS was formed between $z=10.77$ and $z=10.60$ through runaway collisions triggered by rapid star formation. Hereafter, we focus on the $z=10.77$--$10.60$ snapshots to investigate the pollution of the formed SMS.

\begin{figure}
\centering
    \includegraphics[width=0.9\columnwidth]{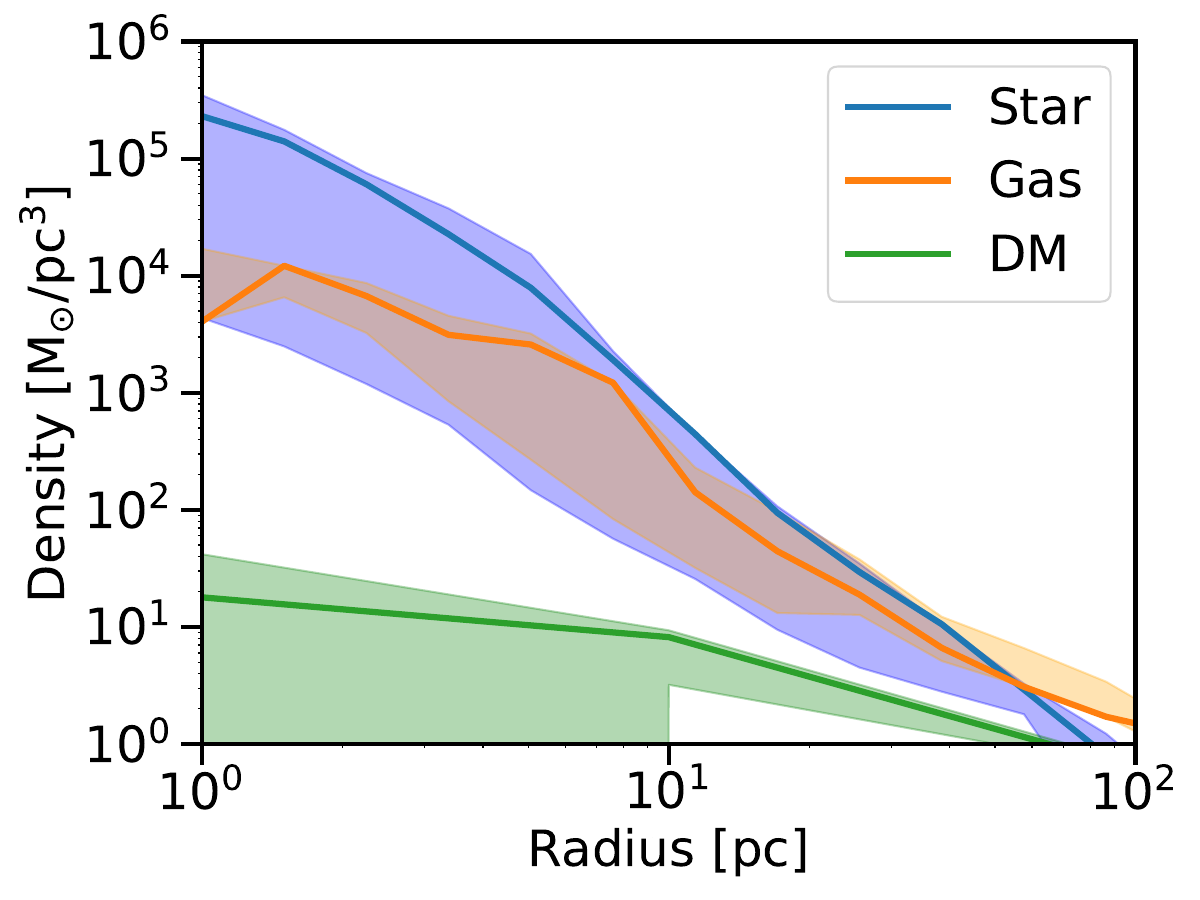}
    \caption{Radial density profiles of dark matter (green), gas (orange), and stars (blue) at $z=10.65$. The shaded regions indicate the possible value ranges of each component during $z=10.77$--$10.60$.
    }\label{fig:303_central_radial_profile}
\end{figure}

In Fig.~\ref{fig:303_central_radial_profile}, we present the density profiles of the gas, star, and dark matter at $z=10.65$, when the SFR reached its peak. For reference, we draw shaded regions for each component from the same density profiles between $z=10.77$ and $ z=10.60$.
At $z=10.77$, the stellar mass was comparable to the gas mass at the galactic center, but at $z=10.65$ and later, stars were dominant compared to the gas. This is because accreting gas was consumed by rapid star formation. In the central region, baryons dominate dark matter.   

In addition, we calculated the mass-weighted mean age of stars within 10\,pc of the galactic center at $z=10.60$, which was 5.05 Myr. 
As the lifetime of $>10$\,$\Msun$ stars is set to be $<100$\,Myr in \textsc{CELib}, our galaxy has already been enriched by WR at this moment, but the pollution from AGB stars has not been activated yet.

Assuming the observed lines arise from the high-temperature gas ionized by the SMS at the galactic center, we focus on the chemical abundances of the gas within 10\,pc of the galactic center. 
As summarized in Table~\ref{tab:abundance_examples}, we measured the abundance ratios of H, C, N, O, and He. The abundance ratios of C, N, and O were $\log({\rm N/O})=-1.81$--$-0.90$ and $\log({\rm C/O})=-0.65$--$-0.05$ for $z=10.77$--$10.60$. 
On the other hand, O/H and He/H changed little in the same time range, $12+\log({\rm O/H})=7.83$--$7.86$, and ${\rm He/H}=0.083$--$0.089$.
Since we focused on chemical abundances within an ionized region illuminated by the SMS's radiation, we assumed a 10\,pc radius, typical of an H\textsc{ii} region.
Note that if we consider the dilution within a radius of 30 \, pc, $\log({\rm N/O})$ changes by at most 0.02 dex. Thus, the results are not affected by the choice of this radius.

In the following, we describe the abundance ratios observed of GN-z11 used for comparison with our simulations.
In this paper, we mainly adopt observed O/H and C/O values for GN-z11 \citep{Isobe2025}, based on the latest JWST data \citep{Bunker2023,Maiolino2024Nature,Maiolino2024,Alvarez2025}, and two assumed electron number densities, $n_\mathrm{e}=10^3$ and $10^5$\,cm$^{-3}$.
For N/O and He/H values that have not yet been reported by \citet{Isobe2025}, we use measurements from other studies \citep{Cameron2023,Ji2025} that assume a low $n_{\mathrm{e}}$ of 100\,cm$^{-3}$.
We confirm that all three studies report/assume consistent electron temperatures and O/H for the low-density case of \citet{Isobe2025}.
We summarize these values in Table~\ref{tab:abundance_examples}.
For comparison, the solar abundance \citep{Asplund2021} is given in the table. 

Without SMS pollution, the N/O was smaller than that of GN-z11 and was even comparable to the solar abundance when $z=10.60$, although the pollution from WR stars (massive rotating stars) has already started in our simulation.
On the other hand, the C/O of our simulation matched the observed value at $z\simeq 10.70$, but exceeded it at later times. 
The O/H ratio in our simulation was close to the observed values, assuming $n_{\rm e}=10^3$\,cm$^{-3}$.

\begin{table*}
 \caption{Chemical abundance N/O-enhanced galaxies.}
 \label{tab:abundance_examples}
 \begin{center}
 \begin{tabular*}{2\columnwidth}{@{}lcccccc@{}}
  \hline
   Name & Redshift & $\log$(N/O) & $\log$(C/O) & 12$+\log$(O/H) & He/H & Reference\\
  \hline
   Our result (without SMS pollution) & $10.77$ & $-1.81$ & $-0.65$ & $7.83$ & 0.083 & -\\[2pt]
    & $10.65$ & $-1.32$ & $-0.38$ & $7.82$ & 0.085 & -\\[2pt]
    & $10.60$ & $-0.90$ & $-0.05$ & $7.86$ & 0.089 & -\\[2pt]
   \hline
   Sun & - &$-0.86$ & $-0.23$ & $8.69$ & 0.082 & (1)\\[2pt]
   \hline
   GN-z11 (Fiducial, $T_\mathrm{e}=1.46\times10^4$ K) & $10.6$ &$> -0.25$ & $> -0.78$ &$7.82\pm0.2$ & - & (2)\\[2pt] 
   GN-z11 ($n_\mathrm{e}=10^3$\,cm$^{-3}$) & $10.6$ & - & $-0.61^{+0.07}_{-0.09}$ & $7.87^{+0.19}_{-0.16}$& - & (3)\\[2pt] 
   GN-z11 ($n_\mathrm{e}=10^5$\,cm$^{-3}$) & $10.6$ & - & $-0.64^{+0.07}_{-0.08}$ & $8.37^{+0.19}_{-0.15}$ &- & (3)\\[2pt]
   GN-z11 & $10.6$ & - & - & - & $0.073^{+0.046}_{-0.029}$ & (4)\\[2pt] 
   \hline
   ID397 & $6.00$ &$-0.67^{+0.14}_{-0.13}$ & - & $7.96^{+0.10}_{-0.08}$ & -  & (5)\\[2pt]
   RXJ2248-ID & $6.11$ &$-0.39^{+0.10}_{-0.08}$ & $-0.83^{+0.11}_{-0.10}$ & $7.43^{+0.17}_{-0.09}$ & $0.166^{+0.018}_{-0.014}$ & (6)\\[2pt]
   GLASS\_150008 & $6.23$ &$-0.40^{+0.05}_{-0.07}$ & $-1.08^{+0.06}_{-0.14}$ & $7.65^{+0.14}_{-0.08}$ & $0.142\pm0.054$ & (7)\\[2pt]
   A1703-zd6 & $7.04$ &$-0.6\pm0.3$ & $-0.74\pm0.18$ & $7.47\pm0.19$ & - & (8)\\[2pt]
   A1703-zd6 & $7.04$ & - & - & - & $0.082\pm0.017$ & (4)\\[2pt]
   GN-z8-LAE & $8.28$ &$-0.44\pm0.36$ & $-0.69\pm0.21$ & $7.85\pm0.17$ & - & (9)\\[2pt]
   CEERS\_01019 (SF) & $8.68$ &$>0.28$ & $<-1.04$ & $7.94^{+0.46}_{-0.31}$ & - & (4)\\[2pt]
   GN-z9p4 & $9.38$ &$-0.59\pm0.24$ & $<-1.18$ & $7.38\pm0.15$ & - & (10)\\[2pt]
   GS-z9-0 & $9.43$ &$-0.93\pm0.37$ & $-0.90\pm0.26$ & $7.49^{+0.11}_{-0.15}$ & - & (11)\\[2pt]
   GHZ9 & $10.15$ &$-0.08$--$0.12$ & $-0.96$--$-0.45$ & $6.69$--$7.69$ & - & (12)\\[2pt]
   GHZ2 & $12.34$ &$-0.25\pm0.05$ & $-0.74\pm0.20$ & - & - & (13)\\[2pt]
   GHZ2 & $12.34$ & - & - & $7.44^{+0.26}_{-0.24}$ & - & (14)\\[2pt]
   MoM-z14 & $14.44$ & $-0.57^{+0.28}_{-0.45}$ & $-0.91^{+0.39}_{-0.22}$ & $7.33^{+0.65}_{-0.56}$ & - & (15)\\[2pt]   
  \hline
 \end{tabular*}
  \tablefoot{Chemical abundance ratios of our simulated galaxy, the sun, GN-z11, and other N/O-enhanced $z\gtrsim6$ galaxies summarized in \citet{Ji2025}.}
 \end{center}
 \begin{flushleft}
 \tablebib{
(1)~\citet{Asplund2021}; (2) \citet{Cameron2023}; (3) \citet{Isobe2025}; (4) \citet{Ji2025};
(5) \citet{Stiavelli2025}; (6) \citet{Topping2024,Yanagisawa2024}; (7) \citet{Isobe2023,Yanagisawa2024}; (8) \citet{Topping2025};
(9) \citet{Navarro-Carrera2024}; (10) \citet{Schaerer2024}; (11) \citet{Curti2025}; (12) \citet{Napolitano2025}; (13) \citet{Castellano2024}; (14) \citet{Calabro2024}; (15) \citet{Naidu2025}.
}
  \end{flushleft}
\end{table*}

\subsection{SMS formation}\label{sec:SMSformation}

Recent numerical simulations suggest that SMSs form via runaway collisions among stars during the formation of star clusters \citep{Fujii2024,Rantala2024}. The central density necessary for runaway collisions is typically $10^6 \Msun$\,pc$^{-3}$. As shown in Fig.~\ref{fig:303_central_radial_profile}, the density of the stellar component is slightly less than $10^6 \Msun$\,pc$^{-3}$. In our simulation, the softening length for baryons is 5.7\,pc, so we cannot resolve the density in the innermost region of this galaxy. If this density profile continues into the inner region beyond 1\,pc, we expect the central stellar density to reach $10^6 \Msun$\,pc$^{-3}$. In the following, we estimate the SMS mass that can be formed in this galaxy. 

From the results of star cluster formation simulations resolving individual stars, \citet{Fujii2024} suggested that 3--5 per cent of the formed stars contributed to the SMSs through runaway collisions; in other words, the accretion rate onto SMSs in star clusters or galaxies is estimated to be 3--5 per cent of the SFR of the host system.
Using this relation, we estimated the SMS growth rate from the SFR of the central region of the galaxy.
In the previous subsection, we measured the SFR of the central region of our simulated galaxy (see Fig.~\ref{fig:long_SFR_connected}). 
We assumed that stars within the central 10\,pc of the galaxy could participate in runaway collisions.

As shown in Fig.~\ref{fig:long_SFR_connected}, the SFR varies with the galaxy's mass accumulation history. It also fluctuates with time. In our simulation, the SFR was always higher than $3\times10^{-1}$\,$\Msun$\,$\mathrm{yr}^{-1}$ after the first peak of star formation at $z\simeq10.77$. On the other hand, the maximum star formation rate achieved during the simulation was 30\,$\Msun$\,$\mathrm{yr}^{-1}$ at $z\simeq 10.67$.
Combining this SFR with the efficiency of the runaway collision, 3--5\, per cent of the formed stellar mass, we can estimate the maximum/minimum accretion rate onto SMS. The lower limit of the accretion rate is calculated to be 3~per cent of $3\times10^{-1}$\,$\Msun$\,$\mathrm{yr}^{-1}$, i.e., $9\times10^{-3}$\,$\Msun$\,$\mathrm{yr}^{-1}$.
Similarly, the maximum accretion rate is calculated to be 5~per cent of 30\,$\Msun$\,$\mathrm{yr}^{-1}$, i.e., 1.5$\Msun$\,$\mathrm{yr}^{-1}$.

Following \citet{Fujii2024} and \citet{2025arXiv250902664P}, we assumed a constant accretion rate. For the mass loss rate, we adopted one depending on the VMS mass ($M_{\rm{VMS}}$) and metallicity ($Z$) \citep{Vink2018}: 
\begin{align}
\label{eq:Vink}
    \log\dot{M}_{\mathrm{VMS}} = -9.13 + 2.1&\log(M_{\mathrm{VMS}}/\Msun)\nonumber\\
    &+0.74 \log(Z/Z_\odot)\ \ [\Msun\ \mathrm{yr}^{-1}].
\end{align}
As the metallicity ($Z$) of VMS, we adopted that of the gas within 10\,pc from the galactic center, $1.37\times10^{-3}$ at $z=10.65$, which is $Z\simeq0.1$\,$Z_\odot$ assuming the solar metallicity of $Z_\odot=0.0134$ \citep{Asplund2009}. 

Assuming a constant accretion rate, the mass loss and accretion rates balance at a certain VMS mass because the mass loss rate increases with VMS mass. With the minimum accretion rate ($9\times10^{-3}$\,$\Msun$\,yr$^{-1}$), the SMS mass converged to 5.0$\times 10^3 \Msun$ (see Fig.~\ref{fig:calc_massloss_min}). 
In this case, the formation timescale was $\sim 1$\,Myr, which is sufficiently shorter than the lifetime of SMSs (see Fig.~\ref{fig:T_eff_evolution}).
With the maximum accretion rate ($1.5$\,$\Msun$\,yr$^{-1}$), in contrast, the SMS mass converged to 5.7$\times 10^4 \Msun$ (see Fig.~\ref{fig:calc_massloss_max}), and the formation time was only 0.1\,Myr.
Thus, we estimated $5\times10^3$--$6\times 10^4$\,$\Msun$ as the SMS mass range that can be formed at the galactic center.

\subsection{Abundance pattern with SMS pollution} \label{sec:Mixingejecta}
\begin{figure*}
    \centering
    \includegraphics[width=0.74\columnwidth]{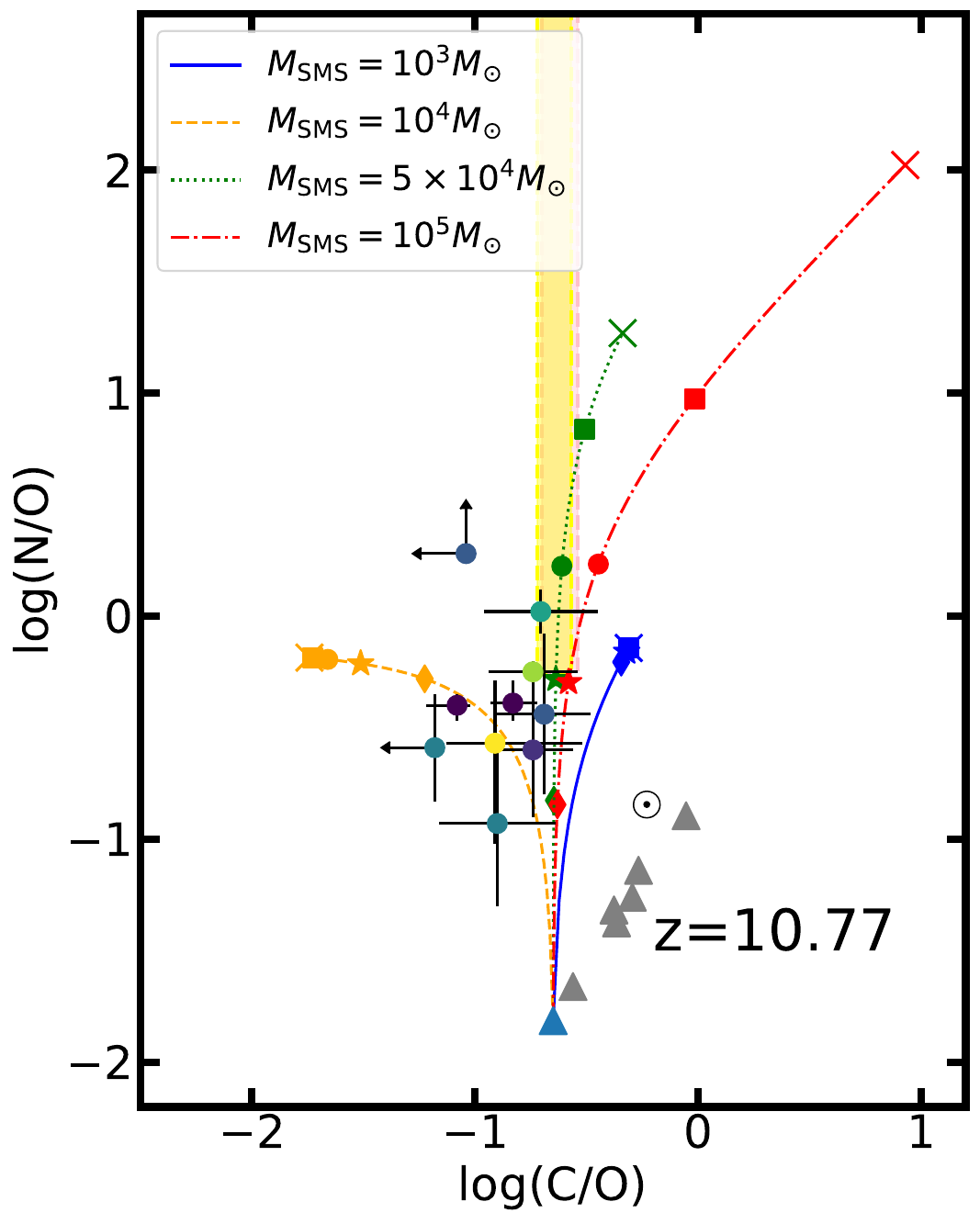}
    \hspace{0.5cm}
    \includegraphics[width=1.06\columnwidth]{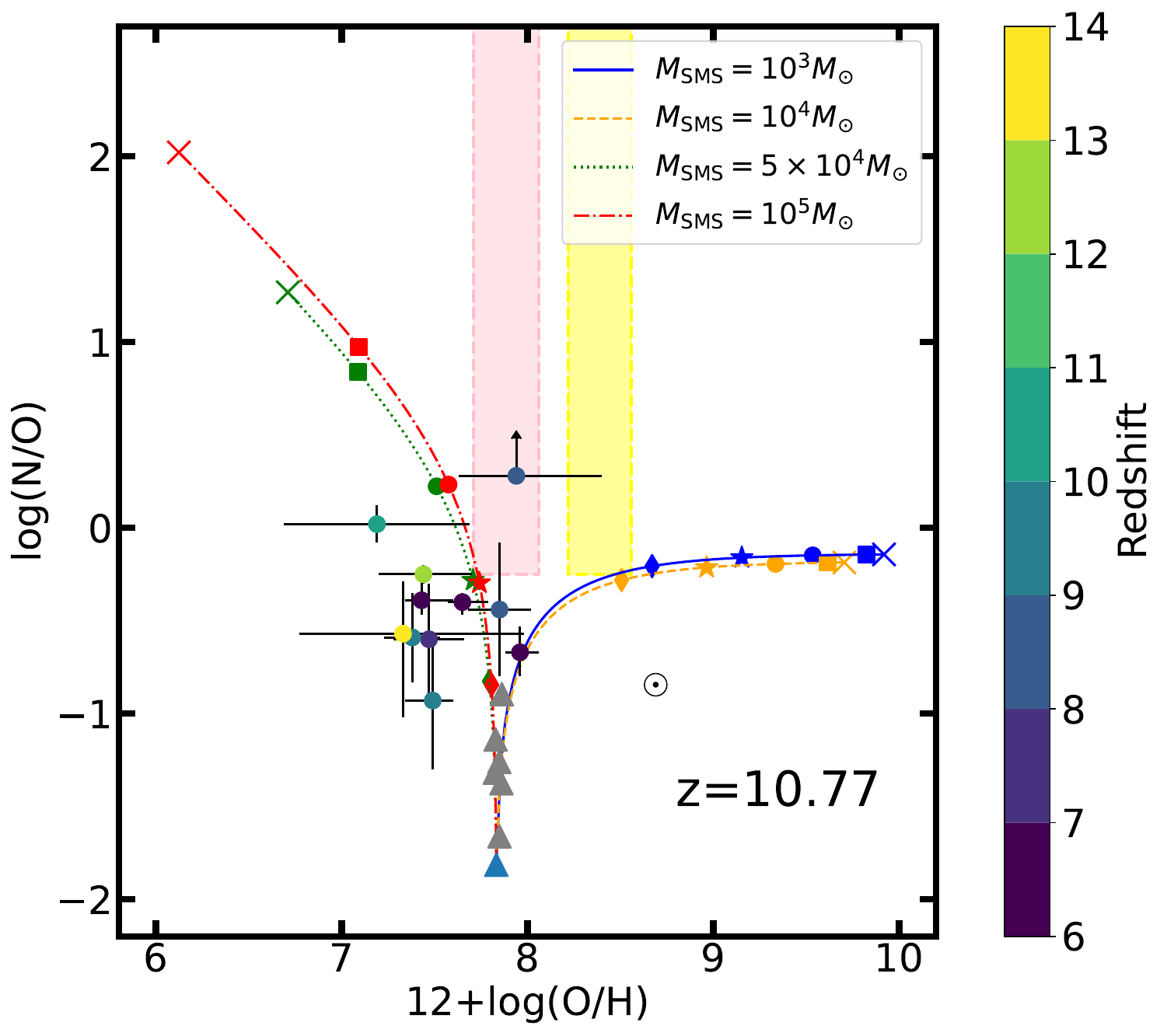}
    \includegraphics[width=0.74\columnwidth]{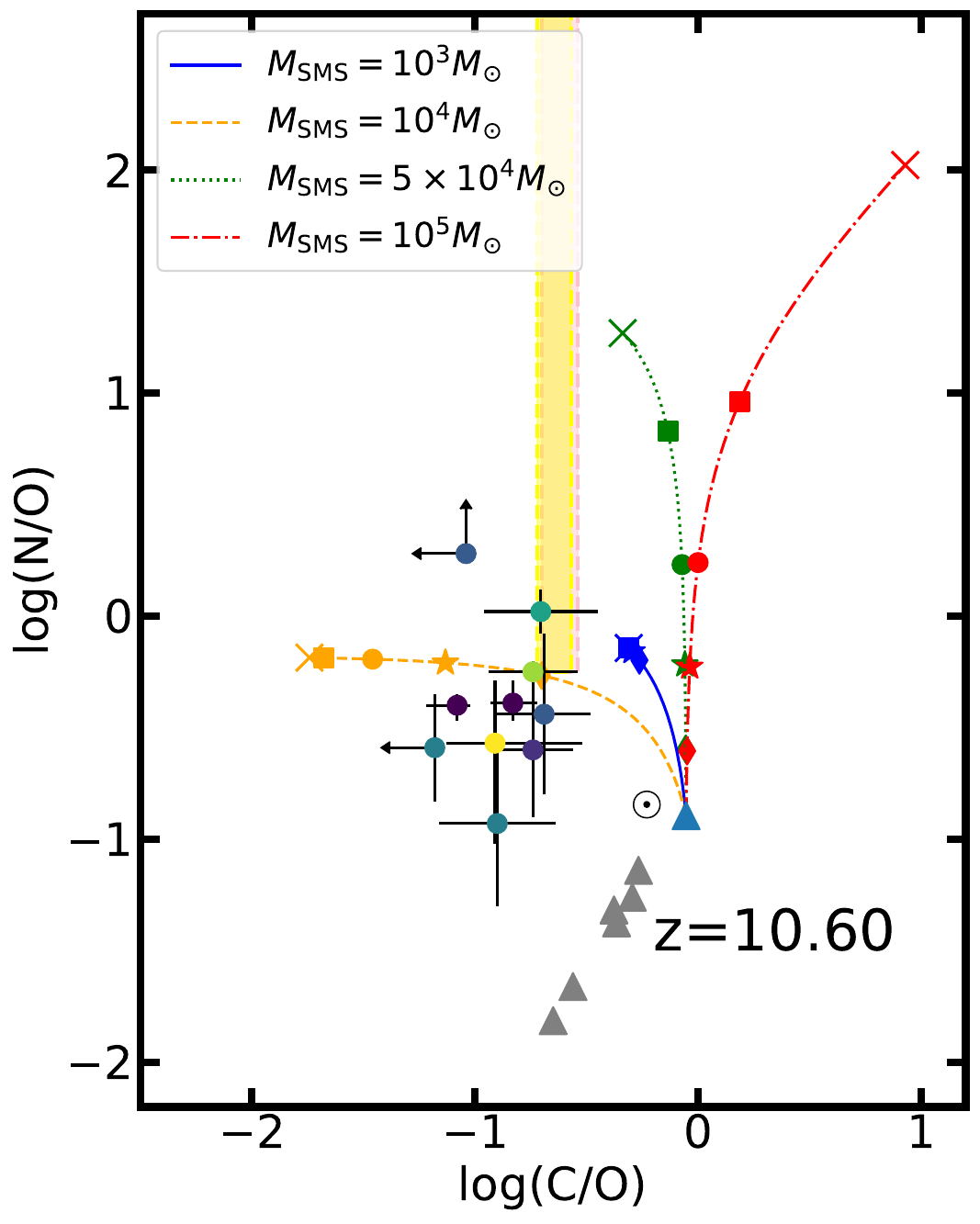}
    \hspace{0.5cm}
    \includegraphics[width=1.06\columnwidth]{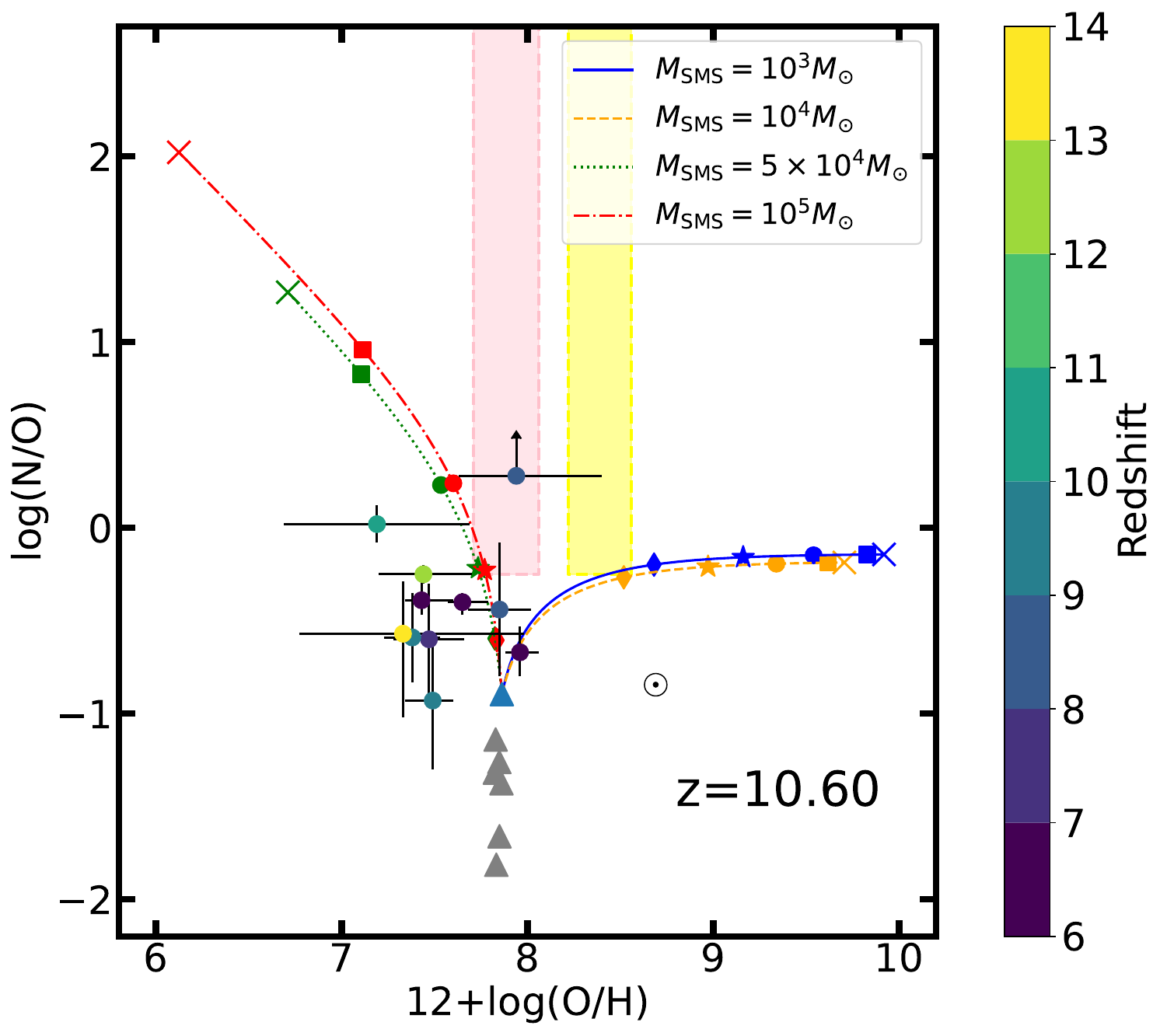}
    \caption{
    $\log({\rm N/O})$ as a function of $\log({\rm C/O})$ (left) and $12 + \log({\rm O/H})$ (right).
    The gray triangles indicate the chemical abundance of our simulated galaxy without SMS pollution between $z=10.77$ and $10.60$. In the top two panels, the blue triangle indicates $z=10.77$. In contrast, in the bottom two panels, the blue triangle represents $z=10.60$. We added SMS ejecta for these two different times. 
    The blue solid, yellow dashed, green dotted, and red dash-dotted indicate the abundance ratio for $10^3$\,$\Msun$, $10^4$\,$\Msun$, $5\times10^4$\,$\Msun$, and $10^5$\,$\Msun$ SMSs, respectively. 
    The symbols (rhombus, star, circle, square, and cross) indicate the pollution fraction ($f_\mathrm{SMS}$) of 10, 30, 60, 90, and 100\, per cent, respectively. 
    The pink shaded region indicates the chemical abundance of GN-z11, the combined log(N/O) value of \citet{Cameron2023}, and the log(C/O) and log(O/H) values from \citet{Isobe2025} with an electron number density $n_\mathrm{e}=10^3$\,cm$^{-3}$. The yellow shaded region represents the GN-z11's abundance from log(N/O) in \citet{Cameron2023} and log(C/O) and log(O/H) from \citet{Isobe2025} with $n_\mathrm{e}=10^5$\,cm$^{-3}$. 
    In addition, the N/O-enhanced galaxies with $z\gtrsim6$ in Table~\ref{tab:abundance_examples} (colored circles with error bars) are plotted. The marker color varies with redshift in the range $6\leqq z\leqq 14$.
    The solar symbol represents the solar abundance \citep{Asplund2021}.}
    \label{fig:chemical_abundance_298_305}
\end{figure*}

In Fig.~\ref{fig:chemical_abundance_298_305}, we present the abundance ratios of C, N, O, and H for the gas within 10\,pc of the galactic center, between $z=10.77$ and $10.60$, obtained from our galaxy formation simulation. 
From $z=10.77$ to $10.60$, both N/O and C/O increased with time, and at $z=10.60$ they reached their maximum values \citep[see also ][]{Saitoh2025}. 

For comparison, we also plotted the abundance patterns of GN-z11 based on observational estimates summarized in Table \ref{tab:abundance_examples}. As mentioned in section \ref{sec:galaxy_formation}, \citet{Cameron2023} adopted an electron temperature of $T_\mathrm{e} =1.46\times10^4$\,K as a fiducial value. \citet{Isobe2025} showed O/H and C/O assuming an electron number density $n_\mathrm{e}=10^3$ cm$^{-3}$ and $10^5$ cm$^{-3}$.
As shown in Fig.~\ref{fig:chemical_abundance_298_305} and Table \ref{tab:abundance_examples}, the N/O ratio of this galaxy is lower than that of GN-z11 without SMSs. 

In the same plot, we also show the abundance pattern obtained by adding the ejecta from each SMS model. 
Similar to \citet{Nagele2023}, we show how the abundance pattern changes as the fraction of the mass from the SMS ejecta varies.
Here, we define the pollution fraction due to SMS ejecta ($f_{\rm SMS}$) as
\begin{equation}
    f_{\rm SMS} = \frac{M_{\rm{ej}}}{M_{\rm{ej}}+M_{\rm gas}},
    \label{eq:mixing_ratio}
\end{equation}
where $M_{\rm{ej}}$ is the ejecta mass of the SMS, and $M_{\rm gas}$ is the gas mass that is in the galaxy and polluted by the SMS ejecta.
As shown in the figure, the $5\times 10^4$ and $10^5\Msun$ SMS models at $z=10.77$ can reproduce the observed abundance pattern of GN-z11 when $f_{\rm SMS}\simeq 30$\,per cent. The observed O/H depends on the assumed electron number density. For the $5\times 10^4$ and $10^5\Msun$ models, $n_{\rm e}=10^3$\,cm$^{-3}$ was consistent with the observation. On the other hand, the $10^4\Msun$ model at $z=10.60$ can also reach the abundance pattern of GN-z11, when $f_{\rm SMS}\simeq10$\,per cent. The O/H value mixed with the $10^4\Msun$ SMS model matched the O/H obtained with $n_{\rm e}=10^5$\,cm$^{-3}$. 

We now estimate $f_{\rm SMS}$ assuming that (1) the radiation/explosion from the SMS forms an H\textsc{ii} region and (2) the ejecta from the SMS pollute the gas in it. 
We assume that the SMS ionizes the surrounding gas and forms an H\textsc{ii} region, and the gas in the H\textsc{ii} region is fully mixed with the SMS ejecta. 
As mentioned above, we estimated that the SMS mass in this galaxy is $5\times 10^3$--$6\times10^4\Msun$. Because the $10^3$, $10^4$, and $5\times10^4\Msun$ SMS models do not explode as supernovae (see Table~\ref{tab:Nagele_2023_table}), we estimate the size and mass of the H\textsc{ii} region using the stellar temperature and radius of the $10^4\Msun$ SMS model (see Sec.~\ref{sec:SMS_models}).
We assumed that these SMSs formed the Str\"omgren sphere under a uniform gas density at the galactic center.
In real galaxies, however, gas distribution is hardly uniform. As shown in Fig.~\ref{fig:303_central_radial_profile}, the gas distribution in the central region is not uniform on a 10\,pc scale. In general, the geometry of ionized regions is non-spherical.
If the gas density at the center of GN-z11 is not uniform, the Str\"omgren radius can lengthen in a direction where the gas is optically thin.
Considering the inhomogeneity of the gas distribution, the $f_{\rm SMS}$ could also vary from a few to a few 10\,per cent.
Observationally, \citet{Alvarez2025} suggested that the ionized radius of GN-z11 is $\simeq64$ pc, and \citet{Xu2024} the estimated CIII] emission's half-light radius is $\sim300$ pc,
although the observational resolution is not sufficiently high to resolve the 10\,pc scale.

For the size of the H\textsc{ii} region, we adopt the Str\"omgren radius \citep[$R_{\rm S}$; ][]{Stromgren1939}:
\begin{align}\label{eq:stromgren_radius}
    R_{\rm S}=\left(\frac{3Q}{4\pi n_\mathrm{H}^2\alpha}\right)^{1/3},
\end{align}
where $Q$ is an ionizing photon production rate of the SMS, $n_\mathrm{H}$ is a number density of neutral hydrogen atoms, and $\alpha$ is a recombination coefficient \citep{2011ApJ...732..100D}:
\begin{align*}
    \alpha = 2.56\times10^{-13}\times(T/10^4\ \mathrm{K})^{-0.83}\ [\mathrm{cm}^3\ \mathrm{s}^{-1}].
\end{align*}
To calculate the Str\"omgren radius, we need $n_{\rm H}$ and $Q$. One may think that we can obtain $n_{\rm H}$ from the simulation, but the softening length for the baryons in our simulation is $\sim 6$ pc, so that we cannot resolve the gas distribution in the innermost region. 
We, therefore, assume a uniform density of the gas; $n_\mathrm{H}=10^3$, $10^4$, and $10^5$\,$\mathrm{cm}^{-3}$, considering uncertainties. As shown in Fig.~\ref{fig:303_central_radial_profile}, the central gas density in our simulation was $\sim 10^4$\,$\Msun$\,$\mathrm{pc}^{-3}$, i.e., $\sim 10^5$\,$\mathrm{cm}^{-3}$. 
On the other hand, recent JWST observations suggested that electron densities of $10^3$\,$\mathrm{cm}^{-3}$ for galaxies at $z\sim10$ \citep{Abdurro'uf2024}. We therefore set these values as higher and lower limits.

We estimate $Q$ using the effective temperature obtained from the $10^4\Msun$ SMS model; $T_{\mathrm{eff}} = 7.09\times10^4$\,K (the maximum value) (see Section \ref{sec:SMS_models}).
The value of $Q$ was calculated with
\begin{align*}
    Q=4\pi R^2 \int\int_{h\nu=13.6\mathrm{\ ev}}^\infty\frac{B(\nu,T_{\mathrm{eff}})}{h\nu}\cos\theta d\nu d\Omega,
\end{align*}
where $R$ is the radius of the SMS, $T_{\mathrm{eff}}$ is the effective temperature, and $B(\nu, T_{\mathrm{eff}})$ is Planck's law \citep{Planck1901}. 
Using the maximum effective temperature ($T_{\mathrm{eff}}=7.09\times10^4$\,K) and the radius at this moment ($R=124$\,$R_\odot$), we obtained $Q=2.68\times10^{52}$\,$\mathrm{s}^{-1}$. 
The resultant Str\"omgren radius, $R_{\rm S}$, and the mass within the radius, $M_{\rm S}$, are 
\begin{equation}
R_{\rm S} = 1.63\times10^3\times \qty (\frac{n_\mathrm{H}}{1\,\mathrm{cm}^{-3}})^{-2/3}\ [\mathrm{pc}],
\end{equation}
and 
\begin{equation}
M_{\rm S} = 4.59\times10^8\times \qty(\frac{n_\mathrm{H}}{\rm{cm}^{-3}})^{-1}[\Msun],
\end{equation}
respectively.
In Table~\ref{tab:Strömgren_radius}, we summarize $R_{\rm S}$ and $M_{\rm S}$, for each given $n_\mathrm{H}$.
The higher the gas density, the higher the pollution fraction (see Table~\ref{tab:Strömgren_radius}). 
From these results, we found that $f_{\rm SMS}=$10--50\,per cent can be obtained from $n_{\rm H}=10^4$--$10^5$\,cm$^{-3}$. 

\begin{table}
 \caption{The property of Str\"omgren spheres.}
 \label{tab:Strömgren_radius}
 \begin{center}
 \begin{tabular*}{1\columnwidth}{@{\extracolsep{\fill}}cccc@{}}
  \hline
  $n_\mathrm{H}$\,$[\mathrm{cm}^{-3}]$ & $R_{\rm S}$\,$[\mathrm{pc}]$ & $M_{\rm S}$\,$[\Msun]$ & $f_\mathrm{SMS}$\,[per cent]\\
  \hline
  $10^3$ & $16.4$ & $4.47\times10^5$ & $0.97$\\[2pt]
  $10^4$ & $3.50$ & $4.47\times10^4$ & $8.95$\\[2pt]
  $10^5$ & $7.55\times10^{-1}$ & $4.47\times10^3$ & $49.6$\\[2pt]
  \hline
 \end{tabular*}
  \tablefoot{The Str\"omgren radius and gas mass inside the Str\"omgren sphere under the different gas number density. Note that gas density was assumed to be uniform.}
 \end{center}
\end{table}

On the other hand, the $10^5\Msun$ SMS model causes an explosion (see Table~\ref{tab:Nagele_2023_table}). We also estimate the radius of the supernovae shell ($R_{\rm SNR}$) using the estimation given in \citet{Cioffi1988}:
\begin{align}
R_{\mathrm{SNR}}=69.02\times
&\left(\frac{E_{\mathrm{explosion}}}{10^{51}\,\mathrm{erg}}\right)^{31/98}\times
\left(\frac{n_\mathrm{H}}{1\,\mathrm{cm}^{-3}}\right)^{-18/49}\nonumber\\
&\times
\xi_\mathrm{m}^{-5/98}\times\beta^{-3/7}\times
\left(\frac{C_0}{10\ \mathrm{km\ s}^{-1}}\right)^{-3/7}\ [\mathrm{pc}]
,
\end{align}
where $E_{\mathrm{explosion}}$ is an explosion energy, $c_0$ is a sound speed, $\xi_\mathrm{m}$ is a metallicity factor, and $\beta$ is an index. By fixing $c_0=10$\,km\,s$^{-1}$, $\beta=1$ and $\xi_\mathrm{m}=1$, we obtained $R_{\rm SNR}=17.2$--$93.6$\,pc and enclosed gas mass $5\times10^7$--$8\times10^7 \ \Msun$ for $n_\mathrm{H}=10^3$--$10^5$\,cm$^{-3}$. If ejecta were fully mixed with the gas inside the shell, the pollution fraction is only $\sim10^{-1}$ per cent.
In the $10^5\Msun$ SMS case, the shock wave swept up a large amount of gas, so pollution cannot be dominant.


\section{Discussion}\label{sec:Discussion}

\subsection{Other high-z N/O-enhanced galaxies}\label{sec:discussion_other_highN_highz}

As well as GN-z11, several other N/O-enhanced galaxies have been found \citep{2025arXiv251120484M}.
We also compared their abundance patterns to our SMS models. For observational data, we used N/O-enhanced galaxies summarized in \citet{Ji2025}, but for $z\gtrsim 6$, which is $\lesssim1$\,Gyr, and therefore we expect less AGB pollution. We also included MoM-z14 \citep{Naidu2025}. 
The results are shown in Fig.~\ref{fig:chemical_abundance_298_305}.
The N/O is mostly higher than the solar abundance, and three of ten have N/O as high as that of GN-z11. The C/O of these galaxies is lower than the solar abundance and, in most cases, lower than that of GN-z11. Their O/H is also subsolar and comparable to or lower than that of GN-z11. There is no clear redshift dependence.

For their C/O values even lower than that of GN-z11, our $10^4\Msun$ SMS model that produces lower C/O or massive SMS models ($5\times 10^4$ and $10^5\Msun$) with low C/O of galactic gas ($z=10.77$) is preferable (see the left panels of Fig.~\ref{fig:chemical_abundance_298_305}). For their O/H values lower than that of GN-z11, our $5\times 10^4$ and $10^5\Msun$ SMS models are preferable (see the right panels of Fig.~\ref{fig:chemical_abundance_298_305}). However, a lower O/H can be realized with a slightly lower O/H of the gas without SMS. 
The O/H ratio here is determined by the gas accreting from the outside of this galaxy, and the O/H ratio reflects the pollution history from Pop-III and II stars and their ejecta. Therefore, the O/H ratio we adopted here ($12+\log\mathrm{(O/H)}=7.82$) may also differ if the galaxy has a slightly different formation history. 
In addition, the abundance pattern of the SMS ejecta does not sequentially change with the SMS mass in our models. 
Thus, other SMS mass models may also explain the abundance patterns of the N/O-enhanced galaxies.

\subsection{Helium abundance}

\begin{figure*}
    \centering
    \includegraphics[width=0.9\columnwidth]{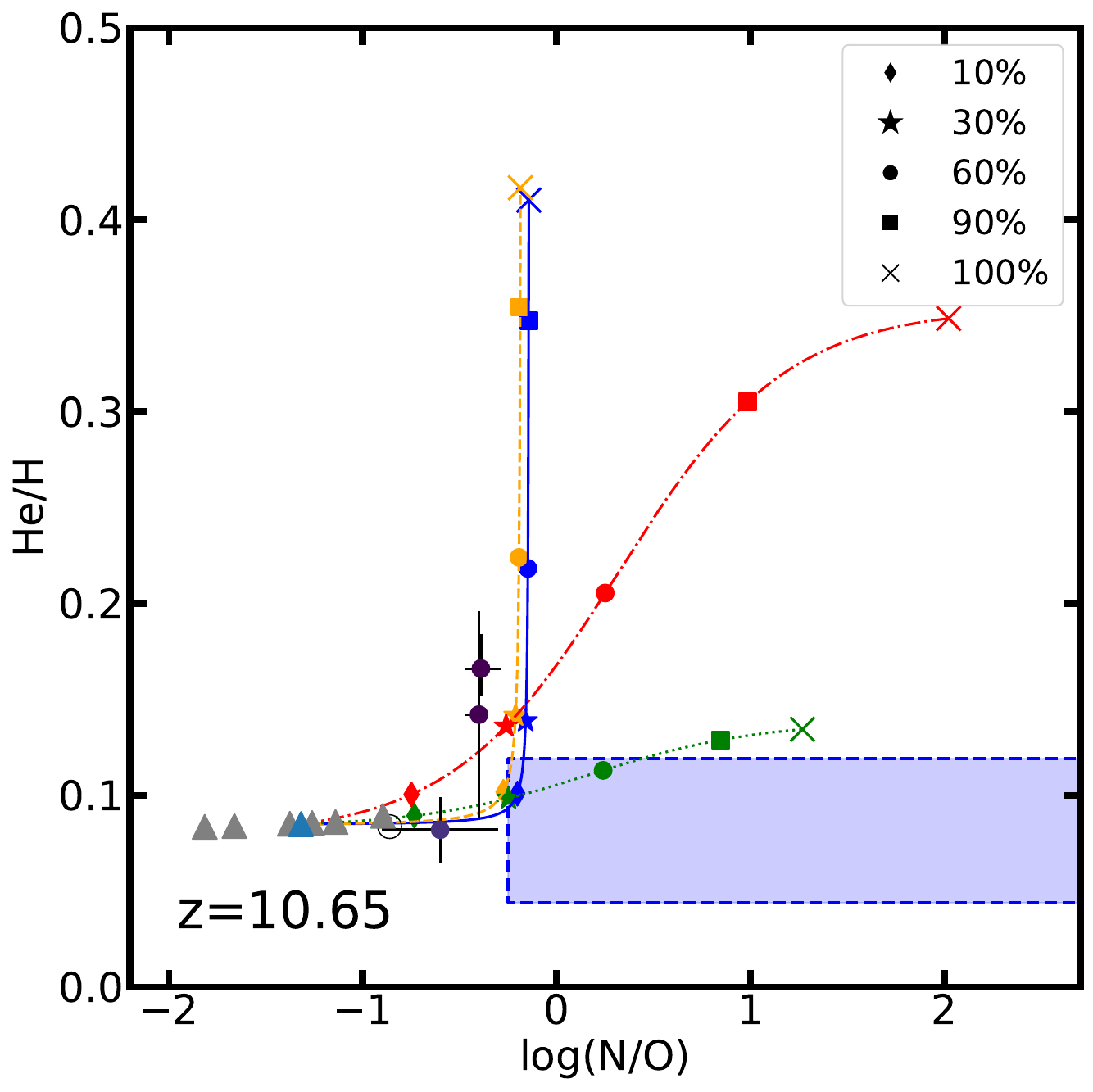}
    \hspace{0.5cm}
    \includegraphics[width=0.9\columnwidth]{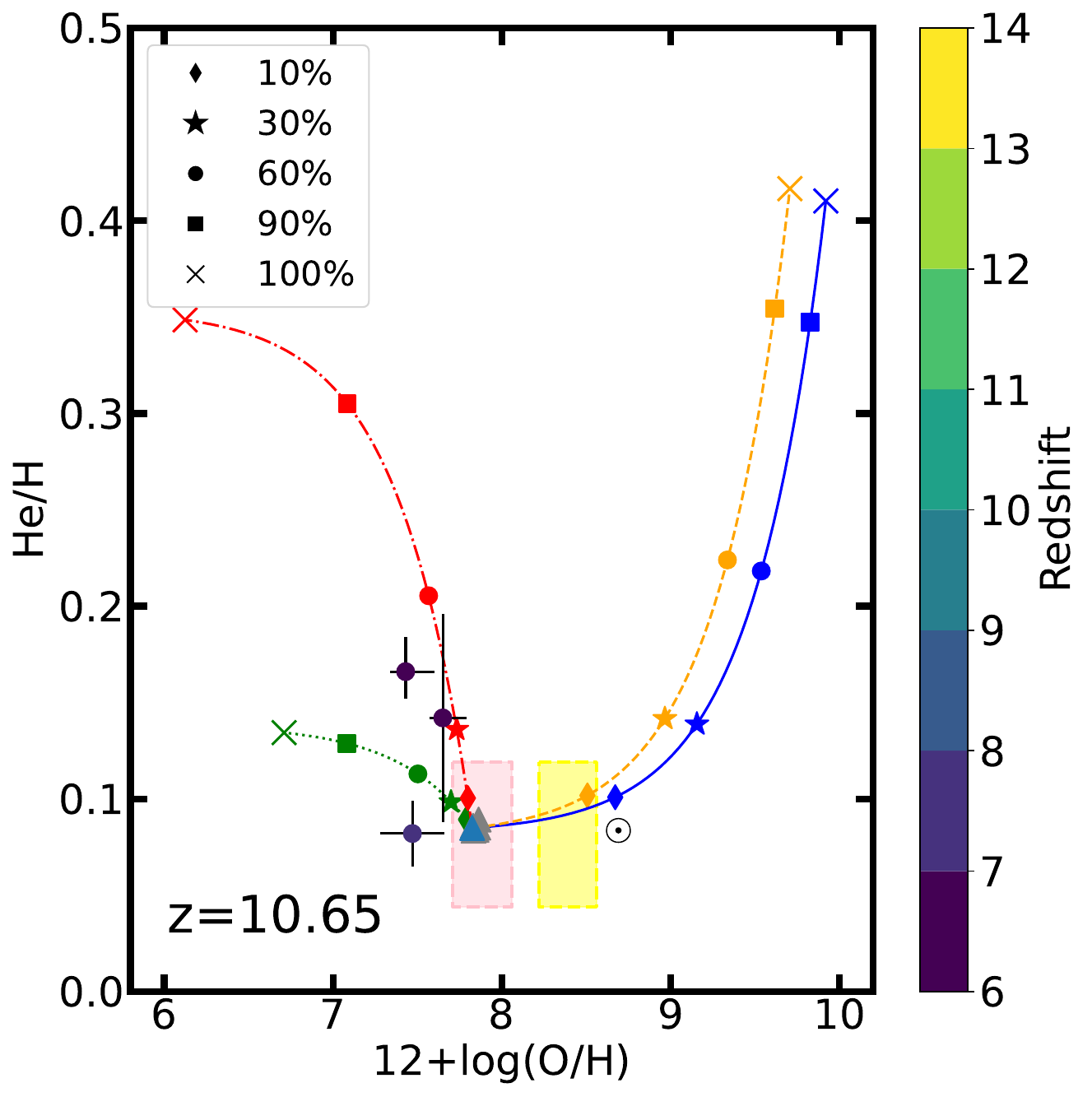}
    \caption{
    Abundance evolution with SMS ejecta pollution. 
    Left: N/O as a function of He/H. Right: O/H as a function of He/H. The initial state at $z = 10.65$ is shown as a blue triangle.
    Colored-circle markers are the same as Fig.~\ref{fig:chemical_abundance_298_305} but only for the N/O-enhanced galaxies for which the He/H are given. 
    The legend of the colored lines, symbols, and gray triangles is the same as Fig.~\ref{fig:chemical_abundance_298_305}.
    The solar abundance is shown by the solar symbol. 
    The blue shaded region indicates the GN-z11's value from \citet{Cameron2023} and \citet{Ji2025}.
    The pink and yellow-shaded regions are also the same as Fig.~\ref{fig:chemical_abundance_298_305}, but we adopted the He/H value  
    from \citet{Ji2025}.
    }
    \label{fig:303_mixed_several_highN_He}
\end{figure*}

Not only C, N, O, and H, but also helium (He) abundances of N/O-enhanced galaxies, including GN-z11, have been measured \citep{Yanagisawa2024, Ji2025}. In Fig.~\ref{fig:303_mixed_several_highN_He}, we present the helium abundances of the N/O-enhanced galaxies \citep{Ji2025}. 
The helium abundance has been determined in only a few galaxies in the \citet{Ji2025} compilation.

In Fig.~\ref{fig:303_mixed_several_highN_He}, we present the He/H of our models ($z=10.77$--$10.60$) and the observed ones as functions of N/O (left panel) and O/H (right panel), respectively. We added SMS ejecta to our $z=10.65$ result, since the He/H does not change significantly over the $z$ range we adopted (see Table~\ref{tab:abundance_examples}). The He/H in our simulation was consistent with that observed in the N/O-enhanced galaxies and GN-z11 without SMS pollution. 
The He/H increases with SMS pollution but remains low compared to GN-z11 when the pollution fraction is $\lesssim 10$--30\,per cent, depending on the SMS mass. 

For the other galaxies, there are only three N/O-enhanced samples because we limited the redshift to $z>6$. The He/H of them is comparable to or higher than that of GN-z11. Such He/H can be explained by the SMS pollution. 
\citet{Yanagisawa2024} reported a positive correlation between N/O and He/H, and \citet{Ji2025} discussed that there is an anti-correlation between He/H and O/H and that such an anti-correlation can be explained by the SMS scenario. The correlation between N/O and He/H can be reproduced by the SMS pollution. The anti-correlation can also be achieved for massive SMSs ($5\times 10^4$ and $10^5\Msun$).

\section{Summary}\label{sec:Conclusion}

To understand the formation mechanism of N/O-enhanced high-$z$ galaxies such as GN-z11, we performed a post-processing analysis on chemical evolution using a cosmological zoom-in $N$-body/SPH simulation of a GN-z11-like galaxy. Although the simulation includes chemical enrichment due to WR and AGB stars, the N/O ratio of the formed galaxy was lower than that of GN-z11. 
As a post-process, we considered pollution from SMSs. From a previous numerical study on the formation of VMSs/SMSs in forming star clusters \citep{Fujii2024}, we assumed 3--5\, per cent of the forming stellar mass in the central 10\,pc of the galaxy contributes to the SMS mass at the galactic center. With the star formation rate of the central 10\, pc obtained from our simulation, we estimated that an SMS with $10^{3\text{--}5}$\,$\Msun$ can be formed in our simulated galaxy. 

Assuming that all of the ejecta from the SMS are mixed with the gas in the H\textsc{ii} region irradiated by the SMS, we calculated the abundance ratios with SMS pollution.
The high N/O ratio of GN-z11 can be obtained by SMS pollution. 
The abundance pattern changes with the SMS masses \citep{Nagele2023}.
For our simulated galaxy, the yields from the $10^4$--$10^5\ \Msun$ SMS models with the pollution fraction of 10--30\,per cent reproduced the observed abundance ratio of GN-z11. Such a pollution fraction can be achieved assuming the gas with $10^4$--$10^5$\,cm$^{-3}$ ionized by the SMS.  
The He/H obtained from our models is also consistent with that of GN-z11. 

Additionally, we compared our results with other N/O-enhanced galaxies. The N/O-enhanced galaxies have lower C/O ratios than GN-z11; these can be best explained by our $10^4$\,$\Msun$ SMS model. On the other hand, our $\gtrsim5\times10^4$\,$\Msun$ SMS models are preferable for the low O/H of the N/O-enhanced galaxies, but a slightly different galaxy formation history may be able to achieve a lower O/H or lower C/O than our current galaxy model.
Similarly to N/O-enhanced galaxies, GCs exhibit N/O-enhanced features \citep{Charbonnel2023,Senchyna2024,Ji2025}. 
Since VMS formation through runaway collisions with stars \citep{Fujii2024} and pollution can also operate in globular clusters, SMS’s pollution scenario can be used to explain the chemical abundances of GCs.
However, not all of the abundance patterns of observed GCs can be explained with VMS/SMSs \citep{2016EAS....80..177C,2018ARA&A..56...83B}.

\begin{acknowledgements}

We thank Chihong Lin and John Silverman for helpful discussions.
The numerical simulation was carried out on the GPU system XD2000 at the Center for Computational Astrophysics (CfCA) at National Astronomical Observatory of Japan (NAOJ). The analysis was also done by the analysis server at CfCA. Sho Ebihara was supported by Forefront Physics and Mathematics Program to Drive Transformation (FoPM), a World-leading Innovative Graduate Study (WINGS) Program, hosted by the University of Tokyo. This study is supported by JSPS KAKENHI Grant No. 24KJ0202, 23K22530, 21K03614, 21K03633, 22KJ0157, 22K03688, 24K07095, 25K01046, and 25H00664.
\end{acknowledgements}

  \bibliographystyle{aa}
  \bibliography{reference}

\begin{appendix}

\section{SMS mass estimation}\label{appendix:SMS_mass_estimation}
\begin{figure}
    \includegraphics[width=\columnwidth]{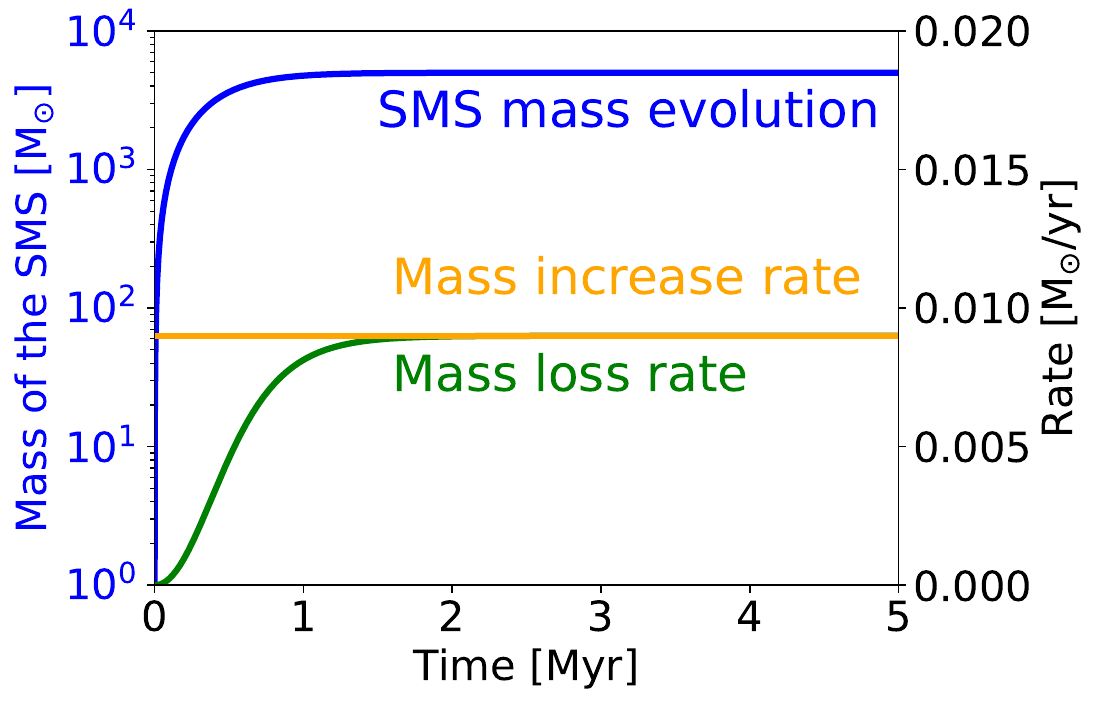}
    \hfill
 \caption{Mass evolution of SMS (blue curve) when an accretion rate was the minimum value,  $9\times10^{-3}$\,$\Msun$\,$\mathrm{yr}^{-1}$ (orange line) and metallicity of 0.1\,$Z_{\odot}$. The green curve indicates the mass loss rate calculated with Eq.~\ref{eq:Vink} \citep{Vink2018}.}
\label{fig:calc_massloss_min}
\end{figure}

\begin{figure}
    \hfill
    \includegraphics[width=\columnwidth]{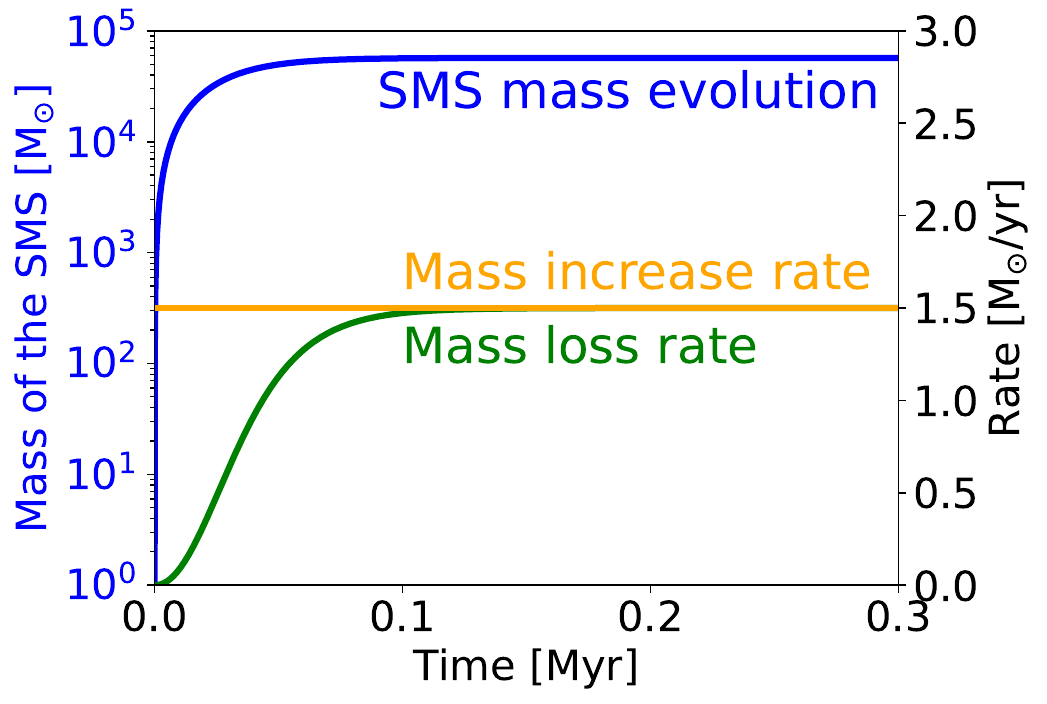}
\caption{
Same as Fig.~\ref{fig:calc_massloss_min}, mass evolution of SMS when the accretion rate was the maximum value, 1.5\,$\Msun$\,$\mathrm{yr}^{-1}$. 
}
\label{fig:calc_massloss_max}
\end{figure}

Although we estimated the SMS mass that can form in the galaxy from the SFR of the galactic center, there would be an uncertainty in the SMS mass.
As discussed in Sec.~\ref{sec:SMSformation}, the accretion rate had an uncertainty of $9\times10^{-3}$\,$\Msun$\,$\mathrm{yr}^{-1}$ to 1.5$\Msun$\,$\mathrm{yr}^{-1}$. Fixing the mass-loss rate of Eq.~\ref{eq:Vink}, we obtained 5.0$\times 10^3 \Msun$ (see Fig.~\ref{fig:calc_massloss_min}). The formation timescale is $\sim 1$\,Myr. Similarly, we also estimated the maximum mass. 
With 5~per cent of 30\,$\Msun$\,$\mathrm{yr}^{-1}$, i.e., 1.5$\Msun$\,$\mathrm{yr}^{-1}$, we estimated 5.7$\times 10^4 \Msun$ as the upper limit of the SMS mass (see Fig.~\ref{fig:calc_massloss_max}). The formation time is only 0.1\,Myr. From these results, we assumed the mass of SMS formed at the galactic center would be in the range of $10^{3-5}$\,$\Msun$.

\section{Variation in the abundance ratios in different snapshots}\label{app:snapshot_variation}

\begin{figure*}
    \centering
    \includegraphics[width=0.9\columnwidth]{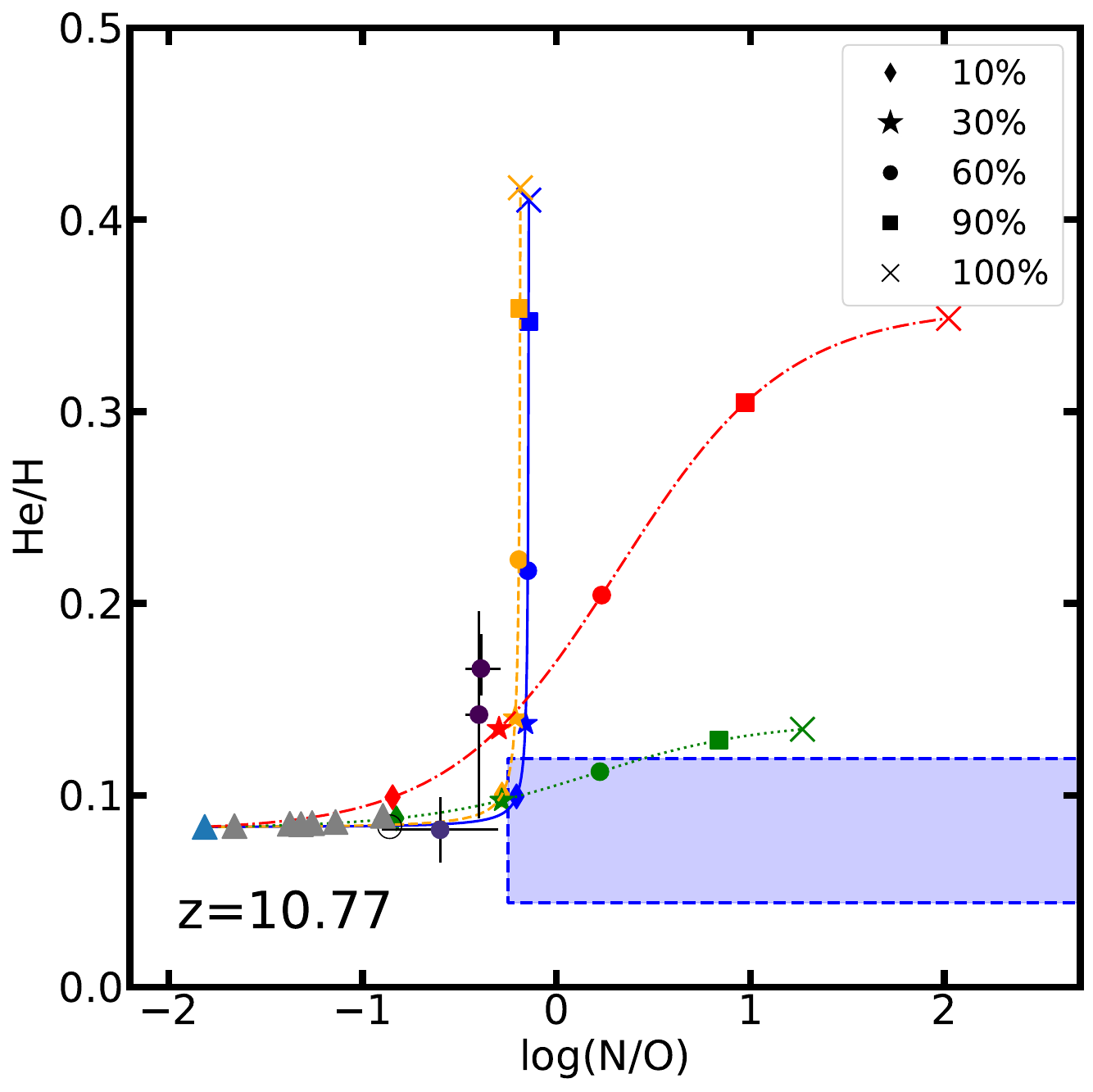}
    \hspace{0.5cm}
    \includegraphics[width=0.9\columnwidth]{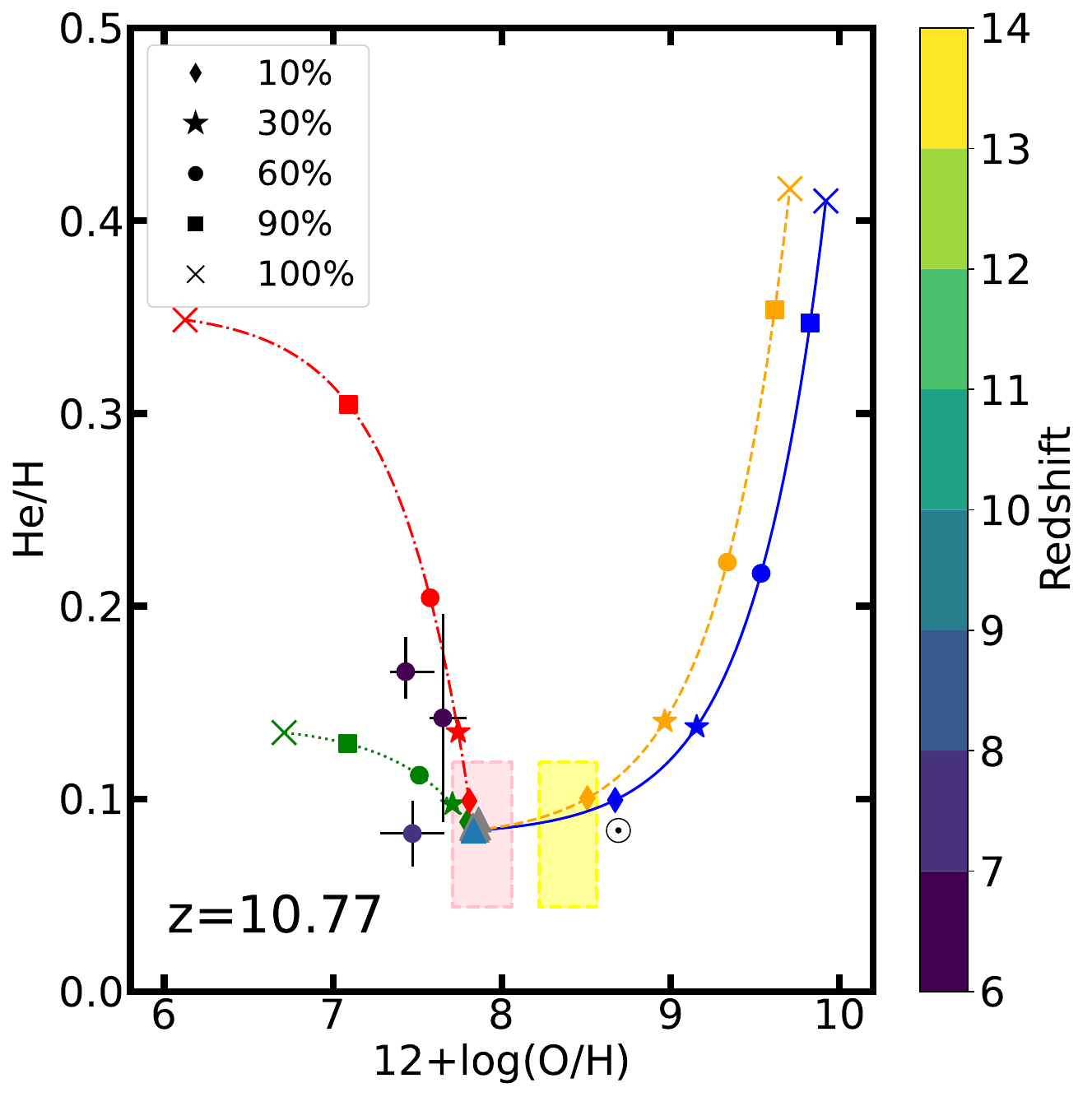}
    \includegraphics[width=0.9\columnwidth]{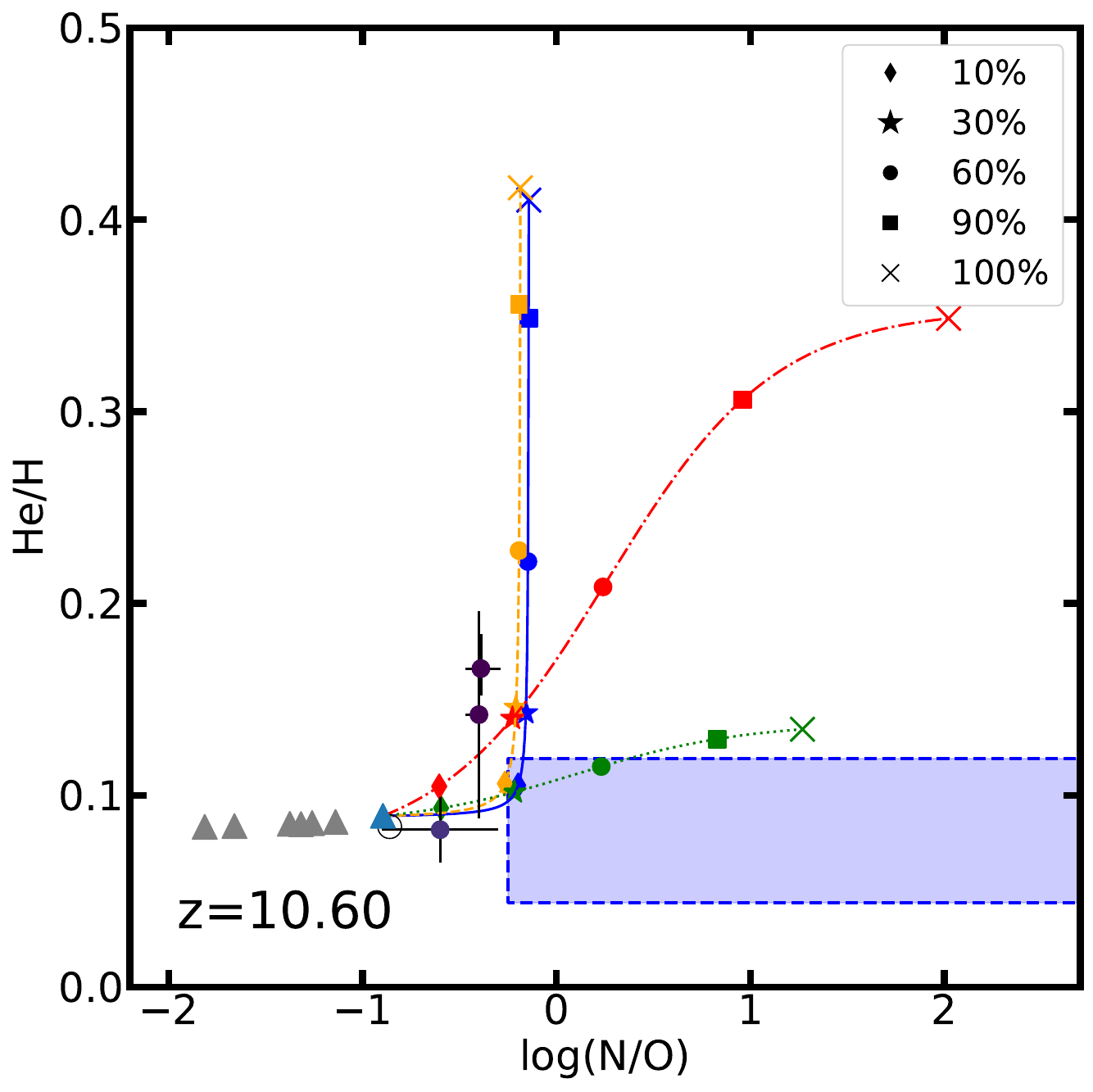}
    \hspace{0.5cm}
    \includegraphics[width=0.9\columnwidth]{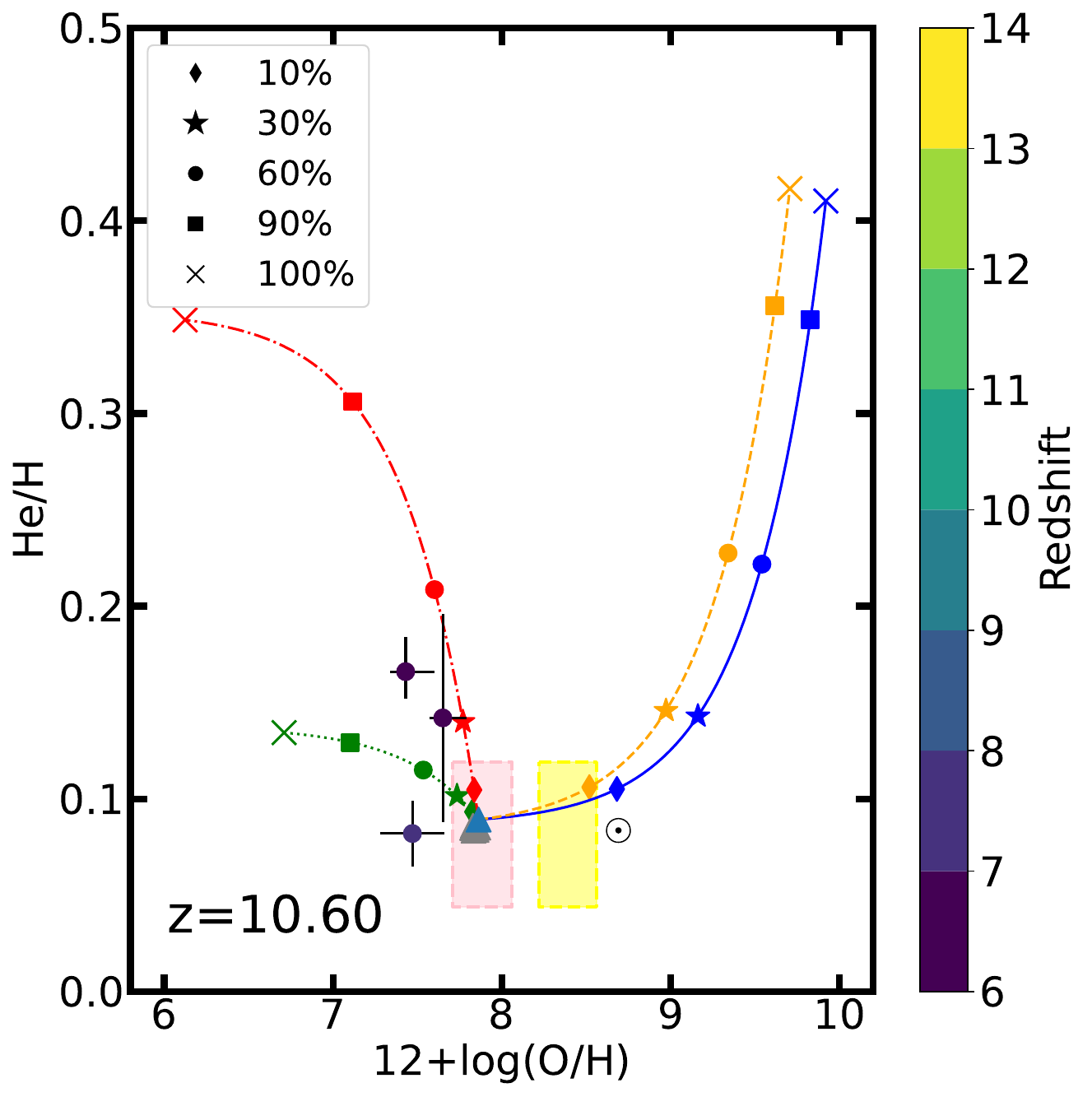}
    \caption{
    As in Fig.~\ref{fig:303_mixed_several_highN_He}, we show N/O vs. He/H and O/H vs. He/H plots for our results and the reported N/O-enhanced galaxies. The results we added SMS ejecta to our $z=10.77$ snapshot (the blue triangle) are shown in the top two panels. The bottom two panels are results for our $z=10.60$ snapshot (the blue triangle). For reference, the chemical abundance at other snapshots, $z=10.77$--$10.60$, is plotted as gray triangles.}
    \label{fig:appendix_He_plot}
\end{figure*}

In the N/O vs. He/H plot and the O/H vs. He/H plot (See Fig.~\ref{fig:303_mixed_several_highN_He}), we added SMS ejecta to our $z=10.65$ result because no matter which snapshot ($z=10.77$--$10.60$) we added the ejecta to, the result does not change.
As proof, we show the N/O vs. He/H plot and the O/H vs. He/H plot in Fig.~\ref{fig:appendix_He_plot}. We added SMS ejecta to our $z=10.77$ result (the top two panels) and our $z=10.60$ result (the bottom two panels). 
Since He/H and O/H do not change significantly, the region SMS model passes by does not change regardless of redshifts.


\FloatBarrier
\clearpage

\end{appendix}
\end{document}